\documentclass[12pt]{iopart}

\usepackage{iopams,graphicx}
\usepackage{lmodern}
\usepackage[T1]{fontenc}
\usepackage{bbold}
\graphicspath{{Graphics/}}
\usepackage{xcolor}

\usepackage{soul}
\usepackage[mathscr]{euscript}
\usepackage{cite}
\usepackage[super]{nth}
\usepackage{float}  
\usepackage{comment}   
\usepackage[colorlinks, citecolor = blue, urlcolor = blue]{hyperref}

\expandafter\let\csname equation*\endcsname\relax
\expandafter\let\csname endequation*\endcsname\relax
\usepackage{amsmath}

\begin{document}

\title[SSD processes with diffusion]{Stochastic synthesis and degradation of arbitrary diffusive processes}

\author{Denis Boyer\footnote{boyer@fisica.unam.mx}}
\address{Instituto de F\'\i sica, Universidad Nacional Aut\'onoma de M\'exico, 
Ciudad de M\'exico 04510, M\'exico}

\author{Gabriel Mercado-V\'asquez\footnote{gabrielmv.fisica@gmail.com}}
\address{Pritzker School of Molecular Engineering, University of Chicago, Chicago, IL 60637, USA}

\begin{abstract}
Target search processes are often carried out by diffusive elements that are introduced and removed at random times, causing fluctuations in the number of searchers in the system over time. For instance, proteins are synthesized and degraded stochastically in the cell, and susceptible to react with specific binding sites during their lifetime. Here we derive the density profile generated by non-interacting particles subject to stochastic synthesis and degradation (SSD) at constant rates, given a source location and an arbitrary propagator for the single, non-degrading particle. We also study the first-passage properties of these many-searcher processes, which can be deduced from the knowledge of the first-passage time distribution of a single particle.
We analyse the behaviours of the characteristic reaction time and of the mean first-passage time in the limits of small and large rates.  Taking into account the costs associated with synthesis leads to the existence of an optimal synthesis rate.
We further generalise these results to bursty synthesis, when production events involve more than one particle at a time, as often observed in cell biology. The mean first-passage time in these situations can be significantly reduced at low synthesis or large degradation rates, as it is divided by the mean burst size in these regimes. The results presented here are applied to one-dimensional Brownian searches. 
Some general connections between SSD and resetting processes are pointed out as well as important differences, since SSD is associated with a distinct queueing protocol.

\end{abstract}

\maketitle

\section{Introduction}

The successful execution of many processes in nature relies on the arrival of particles at a specific domain or target. Therefore, understanding the first-passage-time properties of a system is a ubiquitous objective across physics, chemistry, and biology\cite{redner2001guide,bray_persistence_2013}. Because purely diffusive search can require long times, this problem has motivated a long history of proposed acceleration strategies. Intermittent search strategies—which alternate between slow diffusive exploration and fast ballistic relocation—have been shown to accelerate target arrival across diverse contexts, from animal foraging to intracellular transport \cite{benichou_intermittent_2011,benichou_optimal_2005}. Another search strategy that has gained considerable attention over the past decade is stochastic resetting, in which the searcher's trajectory is interrupted at random times and returned to its starting point \cite{evans_diffusion_2011,evans_diffusion_2011-1}. This simple protocol drives a diffusive particle into a nonequilibrium steady state and, remarkably, minimizes its mean first-passage time (MFPT) to a fixed target at a finite, optimal resetting rate. The idea has since been generalized extensively and reviewed in depth elsewhere \cite{evans_stochastic_2020,evans2014diffusion,pal2017first,gupta2022stochastic}.

However, most treatments of resetting, intermittent search, and the theoretical extensions built around them consider a single entity searching its environment. Although this simplification is useful for tractable mathematical analysis, it remains a significant limitation. Real systems rarely rely on a lone searcher; instead, many searchers act in parallel, and their collective effort can accelerate the overall process. 
At the cellular level, for instance, many copies of the same transcription factor search simultaneously for their specific binding sites on DNA. This collective search accelerates target binding and thereby shapes gene expression\cite{elf2007probing,Pulkkinen_2013}. 
In developmental biology, morphogen gradients emerge from the localized production, diffusion, and degradation of signalling molecules in tissues\cite{Kondo2010_Science,LANDGE20202}. Likewise, synaptic transmission depends on the coordinated release, diffusion, and clearance of many neurotransmitter molecules \cite{RUSAKOV2011359}. 


The environment of living cells is not static but is continuously remodelled through the synthesis, diffusion, and degradation of molecules, and the balance between these processes helps determine reaction kinetics and first-passage statistics. In the immune system, for instance, macrophages respond to pathogens by rapidly increasing the production of cytokines, which diffuse through the extracellular environment until they reach a target cell or are degraded~\cite{Mosser2008-vr,Murray2011-fh,Akira2006-co,Oyler-Yaniv2017-ie,doi:10.1126/scisignal.aaf3998}. Inside cells, the continual production and degradation of molecules is likewise essential for maintaining homeostasis and supporting growth and function. Over the cell cycle, for instance, cells increase protein synthesis to ensure that daughter cells inherit sufficient molecular components~\cite{Alber18042019}.
 
Recently, different approaches have been developed to understand the first-passage times of many-searcher systems \cite{mejia2011first}. The searches carried out by many independent Brownian particles that are launched simultaneously are controlled by the first-passage time of the fastest particle and follow extreme value statistics \cite{lawley2020distribution,lawley2020probabilistic,grebenkov2020single}. The fastest arrivals have also been studied when the search population is continuously increasing, e.g., when particles are injected one after another into the system at a given rate \cite{campos2024dynamics,tung2025first}, or at different times taken from a distribution or a given flux profile \cite{grebenkov2025fastest}. Some predictions have also been made regarding the optimal number of agents that should be used for a given search task, and when to launch them \cite{meyer2025optimal}.
Other approaches focus on “mortal” searchers with finite lifetimes, whose trajectories may terminate before reaching a target site, fundamentally altering first-passage statistics \cite{yuste2013exploration,meerson2015mortality}. 
While redundancy reduces the MFPT, mortality eliminates many particles during the search process. The interplay between these two competing effects has only been recently addressed \cite{SSD_PRE_2026, Linn_PRL_redundancy} and is still not fully understood.

In search processes controlled by stochastic synthesis and degradation (SSD), non-interacting particles are continuously created at a fixed rate $b$ from a localized source, and are independently degraded at a fixed rate $d$ during diffusion \cite{SSD_PRE_2026, Linn_PRL_redundancy}. These systems exhibit four fundamental behaviours \cite{SSD_PRE_2026}. First, the probability that no reaction with a target site has taken place until time $t$ decays exponentially with time and the MFPT is therefore always finite, in contrast with many single-particle search processes. Second, in bounded domains, the MFPT becomes smaller than that of a single searcher with infinite lifetime whenever the synthesis rate
exceeds a threshold which is universal after rescaling, i.e., independent of the type of random motion followed by the individual particles, the spatial dimension or geometry. Third, when one imposes $b=d$, the mean first-passage time for Brownian particles in unbounded space is minimized at a finite rate, in a way similar to optimal resetting. Finally, a biologically motivated energetic cost function determines the synthesis rate 
that 
optimizes the trade-off between search speed and metabolic expense.
 
Here, we extend the theory of SSD processes in several directions. Adopting the point of view of renewal theory, we derive the expression for the position-dependent density of particles, given a point-like source and an arbitrary propagator for the underlying single-particle process. We emphasize some similarities with the position distribution of a particle subject to stochastic resetting, but also some important differences as these two processes are associated to different queueing models. We deduce the non-equilibrium steady state (NESS) that emerges at large times and write down explicitly its form for free Brownian motion in any dimension.


We then focus on the first-passage properties of SSD processes, where the no-reaction probability can be expressed quite generally by knowing the first-passage time distribution of a single, non-degrading particle. We derive the asymptotic behaviours of the MFPT in several limiting cases in terms of the Laplace transform of the above single-particle observable. We use these results to derive an approximate expression of the optimal synthesis rate for Brownian particles in one dimension, when the synthesis costs are taken into account.


Although SSD processes are typically modelled with particles produced one at a time, biological systems often involve the simultaneous production of multiple copies of a molecule. Gene expression is a clear example: mRNA is often produced in bursts, and each mRNA molecule can be translated several times, producing bursts of proteins\cite{Golding2005-realtime,cai2006stochastic,Yu_2006_ProteinBurst,Pedraza_2008,Harper_2011,Kumar_2015}. Similar dynamics arise in viral protein production, where Tat-mediated transcriptional feedback in HIV-1 generates bursty expression that influences the transition between active replication and latency\cite{Weinberger2005-hiv}. Yet first-passage theory for synthesis–degradation processes with non-Poisson statistics remains largely undeveloped. Here, we also extend the SSD framework to incorporate bursty synthesis and study its impact on search efficiency.
 
The remainder of this paper is organized as follows. In section \ref{sec:nb}, we recall the expression for the time-dependent distribution of the number of particles in SSD, as studied in queueing theory. In section \ref{sec:density}, we derive the expression of the density profile and give a simple probabilistic interpretation of this result. The first-passage properties are presented in section \ref{sec:FPT}, while the analysis of the search costs is performed in section \ref{sec:costs}.  Section \ref{sec:bursts} is dedicated to bursty systems. We conclude in section \ref{sec:concl} and some technical details are developed in \ref{sec:appA}.

\section{Number of particles}\label{sec:nb}
\begin{figure}[t]
    \centering
    \includegraphics[width=.8\textwidth]{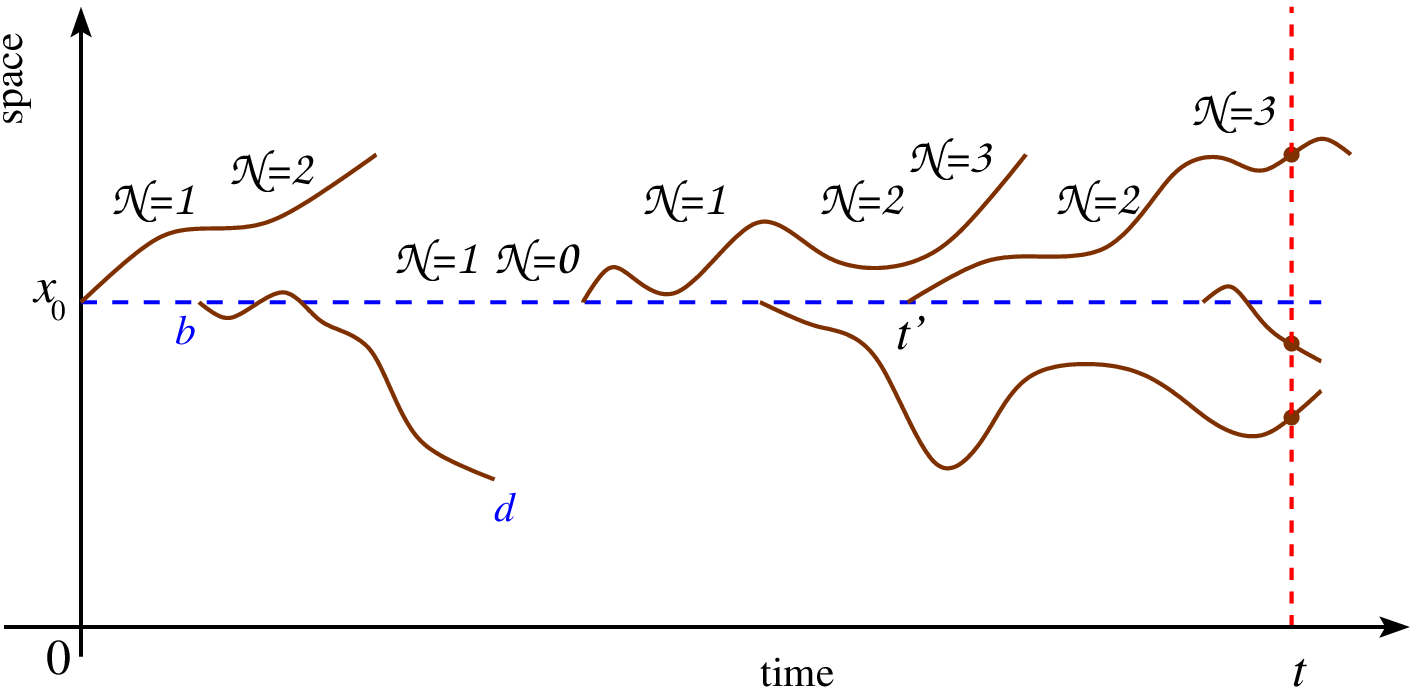}
    \caption{Sketch of an SSD process, where independent particles following an arbitrary stochastic motion are synthesized with rate $b$ at $x_0$ and degrade with rate $d$. Hence the time interval $\tau_s$ between two successive synthesis is exponentially distributed (with pdf $b\,e^{-b\tau_s}$), like the lifetime $\tau_d$ of a particle (of pdf $d\,e^{-d\tau_d}$). The successive numbers of particles ${\cal N}$ in the system are shown for this example. Here, three particles are present at time $t$, with positions $\{x_i(t)\}_{i=1,2,3}$ in a space of arbitrary dimension and geometry.}
    \label{fig:N}
\end{figure}

In the basic SSD model, independent particles of the same type ($A$) are synthesized from a source with constant rate $b$, while the existing particles degrade with a constant rate $d$. The associated reaction is 
\begin{equation}
\emptyset  \overset{b}{\longrightarrow} 
A \overset{d}{\longrightarrow}  \emptyset\, .
\label{reaction}
\end{equation}
Fig. \ref{fig:N} displays schematically a sequence of particle numbers up to some time $t$, given that one initial particle is synthesized at $t=0$.
The number of particles ${\cal N}(t)$ present in the system at time $t$ (irrespective of their positions) is a random variable and we recall here known results about its distribution \cite{gross1985fundamentals}. The mean number of particles in the reaction (\ref{reaction})
satisfies the simple kinetic equation
\begin{equation}
\frac{{\rm d}\langle{\cal N}(t)\rangle}{{\rm d}t}=b-d \langle{\cal N}(t)\rangle\,,
\end{equation}
with solution
\begin{equation}
\langle{\cal N}(t)\rangle=\frac{b}{d}+e^{-dt}\left(n_0-\frac{b}{d}\right)\to \frac{b}{d}\quad {\rm as}\ t\to\infty\, ,
\label{meannb}
\end{equation}
where $n_0$ is the initial number of particles.

Let us denote as $P(n,t)$ the probability that ${\cal N}(t)=n$ with $n\in\{0,1,2,\ldots\}$, given an initial number of particles $n_0$ at $t=0$ (implicit in the notation). 
This process has been mostly studied under the lens of queueing theory. In this context, particles can be seen as "customers" who arrive at some place sequentially and independently at random times (particle birth), are served during a random time (the lifetime of the particle) and then exit the place (particle removal). In Kendall's notation, the SSD process is equivalent to a $M/M/\infty$ queue, meaning that both the arrival and removal of the customers are Markov memoryless processes ($M$), while the number of servers giving service to these customers is infinite: as soon as a new customer arrives it is immediately served, independently of the number of customers being already served. Hence, there is no limit on the number of customers served simultaneously, or the number of particles ${\cal N}(t)$ in the system.

The particle number distribution obeys the master equation
\begin{equation}
\frac{\partial P(n,t)}{\partial t}=-(nd+b)P(n,t)+(n+1)d\,P(n+1,t)+bP(n-1,t)\, .
\label{pkmaster}
\end{equation}
For $n=0$, the last term of Eq. (\ref{pkmaster}) is zero.
Setting the left hand side (lhs) of Eq. (\ref{pkmaster}) to zero yields the steady-state solution $P^{({\rm st})}(n)$, which satisfies $(nd+b)P^{({\rm st})}(n)=(n+1)d\,P^{({\rm st})}(n+1)+bP^{({\rm st})}(n-1)$. It is easy to check that the latter relation is fulfilled by the Poisson distribution with mean $b/d$,
\begin{equation}
P^{({\rm st})}(n)=\frac{(b/d)^n}{n!}\, e^{-b/d}\, .
\end{equation}
The complete time dependent distribution can be derived through the generating function of $P(n,t)$,
\begin{equation}
\hat{P}(z,t)=\sum_{n=0}^{\infty}z^nP(n,t)\,.
\label{generating}
\end{equation}
Applying this transformation to Eq. (\ref{pkmaster}) gives, after re-arranging the terms,
\begin{equation}
\frac{\partial \hat{P}(z,t)}{\partial t}= (z-1)\left[b\hat{P}(z,t)-d\frac{\partial \hat{P}(z,t)}{\partial z}\right]\,.
\end{equation}
Making the change $\hat{P}(z,t)=e^{\frac{b}{d}z}A(z,t)$, one obtains
\begin{equation}
\frac{\partial A(z,t)}{\partial t}+ d(z-1)\frac{\partial A(z,t)}{\partial z}=0\,,
\end{equation}
which can be solved by the method of characteristics. The lines $A(z,t)=cst$ in the $(z,t)$-plane obey the ODE
\begin{equation}
\frac{{\rm d}t}{{\rm d}z}=\frac{1}{(z-1)d}\,,
\end{equation}
or
\begin{equation}
(z-1)e^{-dt}=\xi\,,
\end{equation}
where $\xi$ is an integration constant. Hence $A$ is a function of $\xi$ only, or $A(z,t)=\Phi(\xi)$. Therefore,
\begin{equation}
\hat{P}(z,t)=e^{\frac{b}{d}z}\,\Phi\left((z-1)e^{-dt} \right)\,,
\label{ptildesol}
\end{equation}
while the function $\Phi$ is determined from the initial condition.

\begin{itemize}

\item {\bf Case $n_0=0$}. If no particles are present initially, or $P(n,t=0)=\delta_{n,0}$, then $\hat{P}(z,t=0)=1$. Applying this initial condition to Eq. (\ref{ptildesol}) implies $\Phi(z-1)=e^{-\frac{b}{d}z}$ or $\Phi(\xi)=e^{-\frac{b}{d}(\xi+1)}$. We deduce after substitution back into Eq. (\ref{ptildesol}),
\begin{equation}
\hat{P}(z,t)=e^{(z-1)\frac{b}{d}(1-e^{-dt})}=e^{-\frac{b}{d}(1-e^{-dt})}\sum_{n=0}^{\infty}\frac{z^n}{n!}\left[\frac{b}{d}(1-e^{-dt})\right]^n\,.
\end{equation}
By identification with Eq. (\ref{generating}), the distribution function for this initial condition, denoted as $P^{(0)}(n,t)$, is obtained as a time dependent Poisson distribution,
\begin{equation}
P^{(0)}(n,t)=\frac{K(t)^n}{n!}e^{-K(t)}\quad {\rm with}\ K(t)=\frac{b}{d}(1-e^{-dt})\,.
\label{p0}
\end{equation}

\item {\bf Case $n_0=1$}. Although less often discussed, this is an initial condition of particular interest in the following sections, as search processes commonly start with one particle at $t=0$. In this case, $\hat{P}(z,t=0)=z$ and the distribution, denoted as $P^{(1)}(n,t)$, is obtained similarly by inversion. We present a slightly faster derivation, based on a simple probabilistic argument that will be very useful in the following. An SSD process with $n_0=1$ can be seen as the superposition of two independent processes: the evolution of the initial particle, and that of all the subsequent particles synthesized after $t=0$. The latter is precisely an SSD process with $n_0=0$ initial particles. We can thus write, 
\begin{equation}
P^{(1)}(n,t)=e^{-dt}P^{(0)}(n-1,t)
+(1-e^{-dt})P^{(0)}(n,t)\,,\quad n\ge 1 .
\label{p1p0}
\end{equation}
The first term in the right hand side (rhs) of Eq. (\ref{p1p0}) is the probability that the initial particle has survived up to time $t$ and that, among the particles synthesized afterwards, exactly $n-1$ are present at time $t$. The second term is the probability that the first particle has degraded in the interval $[0,t]$ and that $n$ particles are present (due to the synthesis of subsequent particles). From Eq. (\ref{p0}),
\begin{equation}
P^{(1)}(n,t)=\left[ e^{-dt}+(1-e^{-dt})^2\frac{b}{nd}\right]
\frac{[\frac{b}{d}(1-e^{-dt})]^{n-1}}{(n-1)!}e^{-\frac{b}{d}(1-e^{-dt})}
\,,\quad n\ge 1 \,.
\label{p1p0.b}
\end{equation}

For the special case $n=0$, the first term of the rhs of Eq. (\ref{p1p0}) is absent,
\begin{equation}
P^{(1)}(0,t)=(1-e^{-dt})P^{(0)}(0,t)=(1-e^{-dt})e^{-\frac{b}{d}(1-e^{-dt})}\, .
\end{equation}

\item {\bf Arbitrary $n_0$.} One can repeat the argument used for the case $n_0=1$, which consists of separating the initial particles from the subsequent ones. If exactly $\ell$ initial particles out of $n_0$ have survived at time $t$, a contribution to $P^{(n_0)}(n,t)$ requires $n-\ell$ particles to be present as products of the subsequent synthesis. Therefore, 
\begin{equation}
P^{(n_0)}(n,t)=\sum_{\ell=0}^{n_0}
\binom{n_0}{\ell} e^{-\ell dt}(1-e^{-dt})^{n_0-\ell}\, P^{(0)}(n-\ell,t)\,,
\label{pk0}
\end{equation}
with the convention $P^{(0)}(m,t)=0$ if $m<0$. Alternatively, the sum in Eq. (\ref{pk0}) runs up to $n$ if $n<n_0$.

\end{itemize}

\section{Particle density profiles}\label{sec:density}

\subsection{General formalism}\label{sec:densgen}
Let us now consider a system in arbitrary spatial dimension and with boundaries of arbitrary geometry, where the particles subject to synthesis and degradation (with the rates $b$ and $d$) are synthesized at a position $x_0$ and follow an arbitrary random motion, not necessarily Brownian diffusion. The particles do not interact during their lifetime. In the absence of both degradation and synthesis, the probability distribution function (pdf) of the position $x$ of a single particle starting from $x_0$ at $t=0$ is denoted as $p_{0}(x,t|x_0)$ and is assumed to be known.

An important quantity is the mean particle density at time $t$ of the SSD process, denoted as $\rho_{b,d}(x,t)$, with $b\ge 0$ and $d\ge 0$, where $x_0$ is implicit in the notation for brevity. Formally,
\begin{equation}
\rho_{b,d}(x,t)=\left\langle \sum_{k=1}^{{\cal N}(t)}\delta(x-x_k(t))\right\rangle\,, 
\end{equation}
where $\delta(\cdot)$ is the Dirac delta, $\{x_k(t)\}$ denote the positions of the ${\cal N}(t)$ particles present at time $t$, as exemplified by Fig. \ref{fig:N}, and where the average is taken over all the possible realizations of the SSD process with a given initial condition (by convention, the sum is zero if ${\cal N}(t)=0$). As shown below, the density can be expressed simply as a function of $p_0$,
\begin{equation}
\rho_{b,d}(x,t)=n_0\,e^{-dt} p_{0}(x,t|x_0)+b\int_0^{t}{\rm d}t'\, e^{-d(t-t')}p_{0}(x,t-t'|x_0)\,,
\label{rhogen}
\end{equation}
where $n_0$ is the initial number of particles. Taking the limit $t\to\infty$ in Eq. (\ref{rhogen}), we see that the density tends to a NESS, i.e.,
$\rho_{b,d}(x,t)\to \rho_{b,d}^{\rm (st)}(x)$ with
\begin{equation}
\rho_{b,d}^{(\rm st)}(x)=b\int_0^{\infty}{\rm d}t\, e^{-dt}p_{0}(x,t|x_0)\,.
\label{rhoness}
\end{equation}

\begin{itemize}
\item[] {\bf Probabilistic interpretation.} Eq. (\ref{rhogen}) can be derived heuristically in a very simple way by exploiting the additivity property of the density (see Fig. \ref{fig:N} with $n_0=1$). The first term in the rhs describes the contribution at time $t$ of the $n_0$ independent particles that start their motion together at $t=0$. Under degradation, the propagator, which is the same for each particle, is weighted by an exponential factor representing the Poisson no-degradation probability since the initial time. One must add to this first contribution that of all the particles synthesized during $(0,t]$, represented by the second term. As illustrated in Fig. \ref{fig:N} for one of the particles, in any time interval $[t',t'+{\rm d}t']$ a new particle is synthesized with probability $b{\rm d}t'$ and this particle can occupy the position $x$ at $t$ with pdf $e^{-d(t-t')}p_{0}(x,t-t'|x_0)$, where $t-t'$ is the time elapsed between the synthesis and observation times. Since all the synthesis events are independent and the density is additive, the integral sums over all the particles that can be produced at different times $t'\in (0,t]$. 
\end{itemize}

\begin{figure}[t]
    \centering
    \includegraphics[width=.47\textwidth]{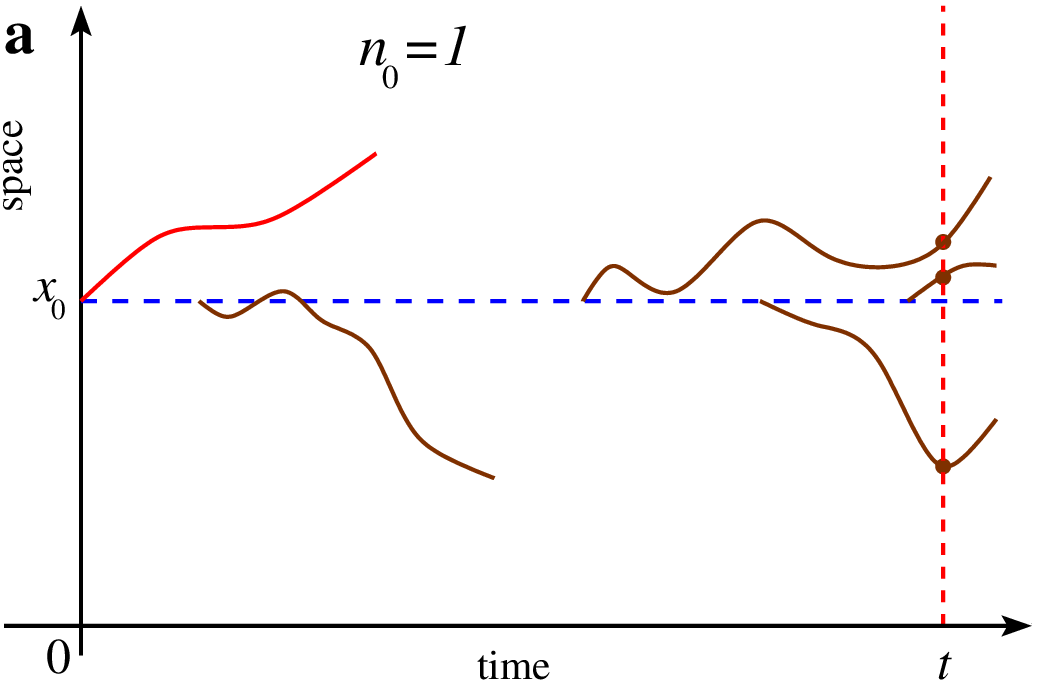}\hspace{0.5cm}
    \includegraphics[width=.47\textwidth]{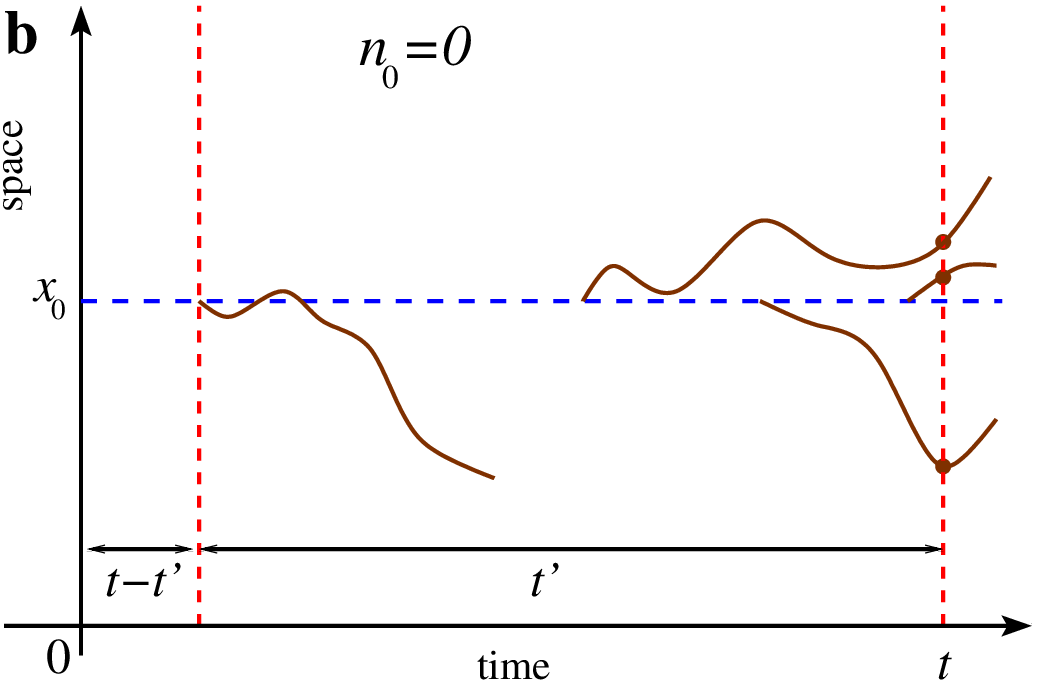}
    \caption{a) An SSD process with $n_0=1$ initial particle can be seen as the superposition of the first particle and the particles produced afterwards, i.e., an SSD process with $n_0=0$. b) The latter is an SSD process with $n_0=1$ but that starts at a later time. It can only contribute to the density at $t$, if the first synthesized particle occurs at time $t-t'\le t$.}
    \label{fig:renew}
\end{figure}

Eq. (\ref{rhogen}) can also be derived in a more pedestrian way by retaking the separation between the $n_0$ initial particles starting from $x_0$ and the following ones, see Fig. \ref{fig:renew}a. For simpler expressions we focus on $n_0=1$, but the generalization is straightforward. Let us denote as $\rho_{b,d}^{(0)}(x,t)$ and  $\rho_{b,d}^{(1)}(x,t)$ the particle densities produced by the SSD processes with $n_0=0$ and $n_0=1$, respectively. We have
\begin{eqnarray}
\rho_{b,d}^{(1)}(x,t)&=&p_{d}(x,t|x_0)+\rho_{b,d}^{(0)}(x,t)\,
\label{rho1rho0}\\
\rho_{b,d}^{(0)}(x,t)&=&b\int_{0}^{t}{\rm d}t'\, e^{-b(t-t')}\rho_{b,d}^{(1)}(x,t')\, .
\label{rho0rho1}
\end{eqnarray}
In Eq. (\ref{rho1rho0}), $p_{d}(x,t|x_0)$ is the propagator of the first particle, subject to degradation. For Poisson degradation, it is simply
\begin{equation}
p_{d}(x,t|x_0)=e^{-dt} p_{0}(x,t|x_0)\,,
\label{p0d}
\end{equation}
as already discussed above.
The second term in the rhs of Eq. (\ref{rho1rho0}) is the contribution from the subsequent particles, i.e., the SSD process in the absence of an initial particle (brown trajectories in Fig. \ref{fig:renew}a). With $n_0=0$, Eq. (\ref{rho0rho1}) translates the fact that only the particles synthesized during $(0,t]$ contribute to $\rho^{(0)}_{b,d}(x,t)$: the factor $e^{-b(t-t')}$ is the probability that no particle is produced in $[0,t-t']$ (with $0<t'\le t$), and $b{\rm d}t'$ is the probability that the first particle appears at $t-t'$. In this case, the time $t-t'$ represents the beginning of an SSD process with $1$ initial particle, whose density evolves during the remaining $t'$ time units until reaching the present time $t$ (Fig. \ref{fig:renew}b). 

Eq. (\ref{rho0rho1}) has a structure somehow similar to a renewal equation and can be solved by introducing $g_{b,d}^{(0)}(x,t)\equiv e^{bt}\rho_{b,d}^{(0)}(x,t)$ and $g_{b,d}^{(1)}(x,t)\equiv e^{bt}\rho_{b,d}^{(1)}(x,t)$. Substituting these expressions, taking the time derivative of Eq. (\ref{rho0rho1}) and eliminating $g_{b,d}^{(0)}(x,t)$ gives,
\begin{equation}
\frac{\partial g_{b,d}^{(1)}(x,t)}{\partial t}=\frac{\partial}{\partial t}\left[e^{bt}p_{d}(x,t|x_0)\right]+b\,g_{b,d}^{(1)}(x,t)\, .
\end{equation}
Solutions are of the form 
\begin{equation}
g_{b,d}^{(1)}(x,t)=a_x(t)\, e^{bt}\,,
\label{gax}
\end{equation}
with 
\begin{eqnarray}
a_x(t)=\int_0^{t}{\rm d}t'\, e^{-bt'}\frac{\partial}{\partial t'}
\left[e^{bt'}p_{d}(x,t'|x_0)\right]+A(x)\,,
\end{eqnarray}
where $A(x)$ is an integration constant independent of time and determined from the initial condition. After integration by parts, $a_x(t)$ reads
\begin{eqnarray}
a_x(t)=p_{d}(x,t|x_0)-\delta(x-x_0)+b\int_0^{t}{\rm d}t'\, p_{d}(x,t'|x_0)+A(x)\,,
\label{ax}
\end{eqnarray}
where we have used the initial condition  $p_{d}(x,t=0|x_0)=\delta(x-x_0)$. From Eq. (\ref{gax}) and the definition of $g_{b,d}^{(1)}(x,t)$, we identify $\rho^{(1)}_{b,d}(x,t)=a_x(t)$. Evaluating Eq. (\ref{ax}) at $t=0$ and using $\rho^{(1)}_{b,d}(x,t=0)=\delta(x-x_0)$ yields $A(x)=\delta(x-x_0)$. Therefore Eq. (\ref{ax}) is equivalent to
\begin{equation}
\rho^{(1)}_{b,d}(x,t)=p_{d}(x,t|x_0)+b\int_0^{t}{\rm d}t'\, p_{d}(x,t'|x_0)\,.
\label{rho1}
\end{equation}
Combined with Eq. (\ref{p0d}), the solution is (\ref{rho1}) the same as Eq. (\ref{rhogen}) with $n_0=1$.

Notice that the mean number of particles in the system $\langle{\cal N}(t)\rangle$ can be re-obtained from the identity $\langle{\cal N}(t)\rangle=\int_{V}{\rm d}x\, \rho_{b,d}(x,t)$ by  integrating Eq. (\ref{rhogen}) over the whole volume. If no absorbing boundaries are present, the propagator is normalized or $\int_{V}{\rm d}x\, p_{0}(x,t|x_0)=1$, and
Eq. (\ref{meannb}) is recovered.

\subsection{Brownian diffusion in $d_{\rm dim}$ dimensions}
In this section, we apply these results to free Brownian particles in an infinite medium of spatial dimension $d_{\text{dim}} = 1, 2, 3$ (not to be confused with the rate $d$). In this case, the propagator is given by the Gaussian distribution in dimension $d_{\text{dim}}$,
\begin{equation}
p_0(\mathbf{x},t|\mathbf{x}_0)=\frac{1}{(4\pi Dt)^{d_{\rm dim}/2}}\, 
\e^{-\frac{|\mathbf{x}-\mathbf{x}_0|^{2}}{4Dt}}\,,
\label{gaussdist}
\end{equation}
where $\mathbf{x}$ is the position vector. We focus here on the non-equilibrium steady state that is reached when $t\to\infty$ and expressed by Eq. (\ref{rhoness}). In the present case,
\begin{equation}
\rho_{b,d}^{(\rm st)}(\mathbf{x})=b\int_0^{\infty}{\rm d}t\, \e^{-dt}\, \frac{1}{(4\pi Dt)^{d_{\rm dim}/2}}\, 
\e^{-\frac{|\mathbf{x}-\mathbf{x}_0|^{2}}{4Dt}}\,.
\label{rhonessgauss.0}
\end{equation}
This type of integral already appeared in Ref. 
\cite{evans2014diffusion} in the context of stochastic resetting. Similarly, we obtain here,
\begin{equation}
\rho_{b,d}^{(\rm st)}(x)=\frac{b}{(2\pi)^{1-\nu}D}\left(\frac{D}{d}\right)^{\nu/2}\,|\mathbf{x}-\mathbf{x}_0|^{\nu}\,K_{\nu}\left(\sqrt{\frac{d}{D}}|\mathbf{x}-\mathbf{x}_0|\right)\,, \quad{\rm with}\ \nu=1-d_{\rm dim}/2\,,
\label{rhonessgauss.1}
\end{equation}
where $K_{\nu}(z)$ is the modified Bessel function of the second kind. In the case $d_{\rm dim}=1$, the identity $K_{1/2}(z)=\sqrt{\pi/(2z)}\,\e^{-z}$ is used, and an exponential particle probability density is obtained,
\begin{equation}
\rho_{b,d}^{(\rm st)}(x)=\frac{b}{2\sqrt{Dd}}\, \e^{-\sqrt{\frac{d}{D}}|x-x_0|}\,,\quad d_{\rm dim}=1\,.
\label{rhonessgauss.2}
\end{equation}
This expression is the same as Eq. (13) in Ref. \cite{evans2014diffusion} if we replace the reset rate $r$ with $d$ and multiply that solution by $b/d$, the average number of particles in the steady state. The other odd dimensions also have closed-form expressions. In the case $d_{\rm dim}=3$, we have $\nu=-1/2$, and since $K_{-1/2}(z)=K_{1/2}(z)$, we deduce
\begin{equation}
\rho_{b,d}^{(\rm st)}(\mathbf{x})=\frac{b}{4\pi D|\mathbf{x}-\mathbf{x}_0|}\, \e^{-\sqrt{\frac{d}{D}}|\mathbf{x}-\mathbf{x}_0|}\,,\quad d_{\rm dim}=3\,.
\label{rhonessgauss.3}
\end{equation}
SSD with Brownian diffusion is a particular case where the density at any given time also satisfies a PDE,
\begin{equation}\label{diff}
\frac{\partial \rho_{b,d}(\mathbf{x},t)}{\partial t}=D\nabla^2\rho_{b,d}(\mathbf{x},t)-d\rho_{b,d}(\mathbf{x},t)+b\delta(\mathbf{x}-\mathbf{x}_0)\, .
\end{equation}
This equation is written by modifying the diffusion equation to incorporate a bulk loss term and a gain term arising from the particle source located at $\mathbf{x}_0$. This equation adequately describes the spatial concentration of fluorescent proteins in many experiments, see e.g. \cite{drocco2012synthesis}. The solution to Eq. (\ref{diff}) for all $t$ is exactly expressed by the more general equation (\ref{rhogen}), where $p_0$ is the free Brownian propagator in Eq. (\ref{gaussdist}). Alternatively, the steady state (\ref{rhonessgauss.1}) can be re-derived from Eq. (\ref{diff}) by setting the time derivative to $0$ and
using the expression for the Laplacian in radial coordinates in $d_{\rm dim}$ dimensions.

We can see that Eq. (\ref{diff}) with $b=d=r$ is identical to the one that describes the probability density of the position of a Brownian particle with stochastic resetting at rate $r$ \cite{evans_diffusion_2011,evans_diffusion_2011-1,evans_stochastic_2020}. In the following section, we discuss the general connection between SSD and resetting processes, beyond the Brownian example.

\subsection{A note on resetting processes}\label{sec:reset}

Let us choose the particular initial condition $n_0=b/d$ (if it is an integer) and define the re-scaled density $\rho^*_{b,d}(x,t)=\rho_{b,d}(x,t)/n_0$. If $b/d$ is not an integer, it can be understood as an ensemble average over different initial particle numbers. In this situation, $\int_V {\rm d}x\, \rho^*_{b,d}(x,t)=1$ at all times, as predicted by Eq (\ref{meannb}). Eq. (\ref{rhogen}) becomes
\begin{eqnarray}
\rho^*_{b,d}(x,t)&=&e^{-dt} p_{0}(x,t|x_0)+d\int_0^{t}{\rm d}t'\, e^{-dt'}p_{0}(x,t'|x_0)
\label{rhotilde}\\
&=&p_{d}^{(SR)}(x,t|x_0)\,.
\end{eqnarray}
The relation (\ref{rhotilde}) for $\rho^*_{b,d}(x,t|x_0)$ is thus identical to $p_{d}^{(SR)}(x,t|x_0)$, which denotes the pdf of the position $x$ of a {\em single} particle starting from $x_0$ at $t=0$, and subject to stochastic resetting (SR) to $x_0$ with rate $d$. Processes under stochastic resetting are defined as follows: during any time interval $[t,t+{\rm d}t]$, with probability $1-d\times{\rm d}t$ the particle updates its position according to its underlying process or propagator; with the complementary probability $d\times{\rm d}t$, the particle is instantaneously brought back to its starting point $x_0$. From there, a new independent path is initiated and proceeds until the next resetting event, and so forth.

One can immediately see that SR and SSD dynamics are very different.
In stochastic resetting, there is only one particle in the system at all times, or $P^{(SR)}(n,t)=\delta_{n,1}$. In other words, synthesis events are perfectly correlated to the degradation events, instead of occurring independently and producing fluctuations in the particle number. In the terminology of queueing theory, SR can be seen as a queue with one server: particles enter the system (the queue) and are served (diffuse) one at a time. As soon as a particle exits the system the next one in the queue is immediately served (starts to diffuse). This additional rule requires that the arrival rate of customers ($\lambda_a$) is very large, so that the number of particles waiting is never zero. In Kendall's notation, Markov resetting processes can be represented as a queue with
\begin{equation}
\underset{\lambda_a\to\infty}{M}/M/1\,,
\end{equation}
which clearly differs from $M/M/\infty$ queues.

Then one may wonder why the densities $\rho^*_{b,d}$ (with the initial number of particles $n_0$ set to $b/d$)
and $p^{(SR)}_d$ both obey Eq. (\ref{rhotilde}).
The answer resides in the different probabilistic interpretations of this equation for the two processes. Instead of representing a sum over different particles as described in section \ref{sec:densgen}, for resetting Eq. (\ref{rhotilde}) is a renewal equation where the second term of the rhs has a different meaning (see, e.g., \cite{evans_stochastic_2020}). The first term in the rhs represents the contribution to the pdf when no resetting event occurs in $[0,t]$ (a situation that has probability $e^{-dt}$), therefore the particle evolves from $x_0$ to $x$ with its propagator
$p_0(x,t|x_0)$. The second term represents the complementary situation when at least one resetting occurs in $[0,t]$. In this case, only the last resetting event before $t$ matters for the probability density at $x$. Let us assume that a resetting event occurs at time $t-t'$ (with probability $d\times{\rm d}t'$, where $d$ is the resetting rate here) and that it is the last one before $t$. The process evolves from there with the propagator $p_0$ from $x_0$ to $x$ during a time $t'$ to reach the present time $t$. The probability that no resetting occurs in the final interval $[t-t',t]$ (so that $t-t'$ is actually the time of the last resetting event) is $e^{-dt'}$. One next multiplies these 3 independent probabilities and sums over all the possible times $t'$. After making the change $t-t'\to t'$, the second term of the rhs of Eq. (\ref{rhotilde}) is obtained.

As shown in the next section, the difference between SR and SSD processes is noticeable in their first passage properties.


\section{First passage times statistics}\label{sec:FPT}

\subsection{General formalism}

\begin{figure}[t]
    \centering
    \includegraphics[width=.8\textwidth]{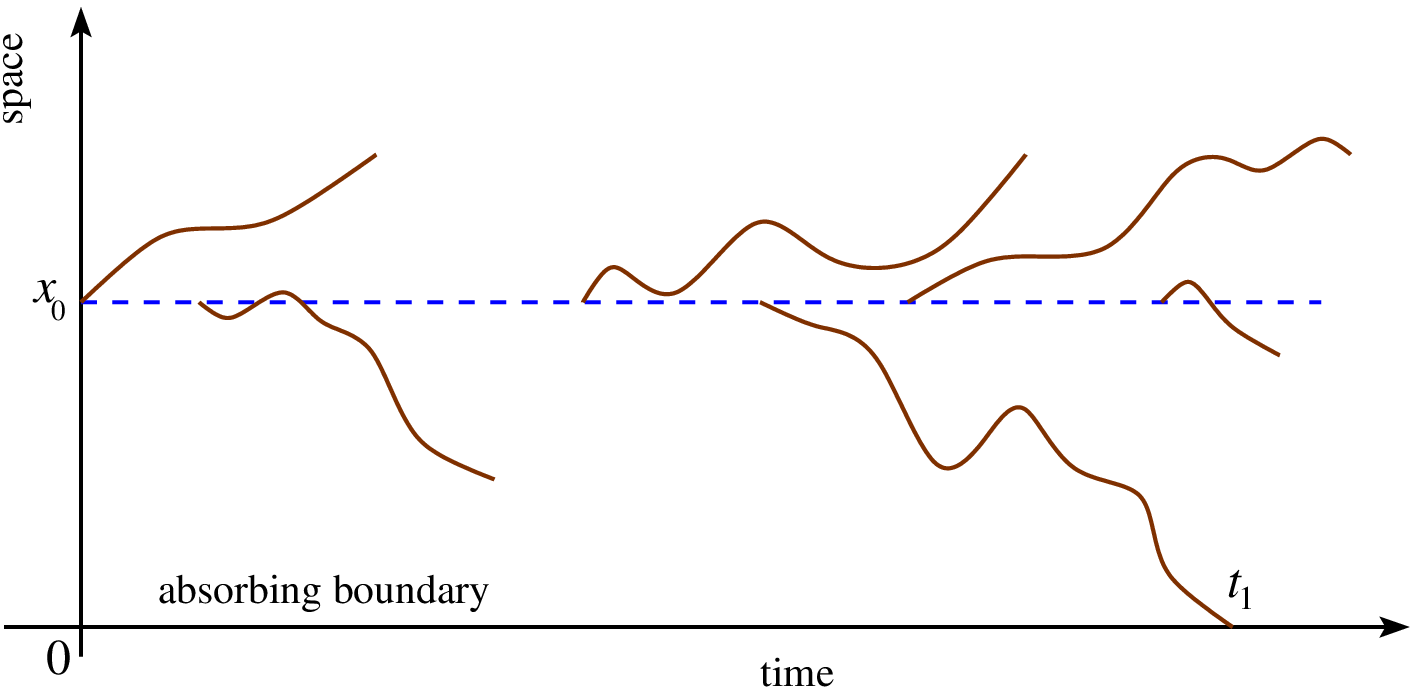}
    \caption{First-passage time $t_1$ of an SSD process with a target region (for instance a point located at the origin in a one-dimensional system).}
    \label{fig:FPT}
\end{figure}

We now investigate search processes carried out by particles subject to synthesis and degradation, given an initial condition at time $t=0$ of $n_0$ particles starting to diffuse from $x_0$. Each particle (initial or produced afterwards) is susceptible to react with a fixed target site or region, for instance the boundary at $x=0$ in a one-dimensional space, see Fig. \ref{fig:FPT}. 
The first-passage time $t_1$ is defined as when the target is first reached by a particle, and we seek to determine its distribution. This problem was studied in \cite{SSD_PRE_2026,Linn_PRL_redundancy} with $n_0=1$ and we follow here a route similar to the one exposed in \cite{SSD_PRE_2026}.

The most basic quantity here is $\varphi_0(t)$, the pdf of the first-passage time at the target site by a single non-degrading particle starting from $x_0$ at $t=0$, which is assumed to be known for the given setup (henceforth, $x_0$ is implicit in the notation). Let us also define $q_d(t)$ as the probability that a single particle starting at $x_0$ at $t=0$ and subject to degradation with rate $d$ has not reacted with the target at time $t$. We have the identity
\begin{equation}
q_d(t)=1-\int_0^{t}{\rm d}\tau\,e^{-d\tau}\varphi_0(\tau)\,.
\label{qdp0}
\end{equation}
Eq. (\ref{qdp0}) expresses the fact that the trajectories that hit the target at a time  $\tau$ less than $t$ (meaning that they have not degraded at time $\tau$, hence the exponential factor in the integral) do not contribute to $q_d(t)$ by definition.

A quantity of primary interest is the no-reaction probability $Q_{b,d}^{(n_0)}(t)$ that the target has not been reached at time $t$ by any particle. We have,
\begin{equation}
Q_{b,d}^{(n_0)}(t)=q_d(t)^{n_0}\, Q_{b,d}^{(0)}(t)\,.
\label{Qbdn0}
\end{equation}
Using again the independence between particles, we have written the total no-reaction probability in Eq. (\ref{Qbdn0}) as the product of the no-reaction probabilities of the $n_0$ initial particles and that of the subsequent particles, which compose an independent SSD process with $0$ initial particles, as sketched in Fig. \ref{fig:renew}a-b. An additional relation links $Q_{b,d}^{(n_0=0)}$ to $Q_{b,d}^{(n_0=1)}$ in a way analogous to Eq. (\ref{rho0rho1}) for the densities,
\begin{equation}\label{Q0Q1}
Q_{b,d}^{(0)}(t)=e^{-bt}+b\int_0^{t}{\rm d} t'\, e^{-b(t-t')} Q_{b,d}^{(1)}(t')\, .
\end{equation}
This is a renewal-like equation, that separates $Q_{b,d}^{(0)}$ into two contributions. The first terms on the rhs of Eq. (\ref{Q0Q1}) is the probability that no particle is synthesized in $[0,t]$: since there are no particles in the system, the no-reaction probability is 1 in this eventuality. The second term considers, as in Eq. (\ref{rho0rho1}), the synthesis of the first particle at a time $t-t'$ with $t'\in[0,t]$, which constitutes the start of an SSD process with $n_0=1$ evolving during $t'$ time-units up to $t$ (see Fig. \ref{fig:renew}b).

Eq. (\ref{Q0Q1}) and Eq. (\ref{Qbdn0}) with $n_0=1$ form a closed system with two unknowns $Q_{b,d}^{(0)}$ and $Q_{b,d}^{(1)}$, which can be solved as \cite{SSD_PRE_2026}
\begin{equation}
Q_{b,d}^{(0)}(t)=\exp\left[-b\int_0^{t}{\rm d}\tau[1-q_d(\tau)]\right]\,,
\label{Q0a}
\end{equation}
to be further simplified after integration by parts, as
\begin{equation}
Q_{b,d}^{(0)}(t)=\exp\left[-b\int_0^{t}{\rm d}\tau\, e^{-d\tau} (t-\tau) \varphi_0(\tau)\right]\,.
\label{Q0b}
\end{equation}
This result was also obtained in \cite{campos2024dynamics,Linn_PRL_redundancy} by other methods.
Inserting Eqs. (\ref{qdp0}) and (\ref{Q0b}) into Eq. (\ref{Qbdn0}) gives the first-passage properties of an arbitrary SSD process in terms of $\varphi_0(t)$, $b$, $d$ and $n_0$. Hence the knowledge of the behaviour of one particle with infinite lifetime determines that of the collection of particles. It is easy to see that $Q_{b,d}^{(0)}$ (and therefore $Q_{b,d}^{(n_0)}$) is an increasing function of $d$ (at fixed $b$) and a decreasing function of $b$ (at fixed $d$). The same properties hold for the MFPTs. 

For $n_0=1$,
\begin{equation}
Q_{b,d}^{(1)}(t)= \left(1-\int_0^{t}{\rm d}\tau\,e^{-d\tau}\varphi_0(\tau)\right)
e^{-b\int_0^{t}{\rm d}\tau\, e^{-d\tau} (t-\tau) \varphi_0(\tau)}\,.
\label{Q1}
\end{equation}
The first-passage time pdf, denoted as $\varphi_{b,d}^{(n_0)}(t)$ where 
$\varphi_{b,d}^{(n_0)}(t){\rm d}t={\rm Prob}[t\le t_1\le t+{\rm d}t]$, follows from the general relation $\varphi_{b,d}^{(n_0)}(t)=-\frac{\partial Q_{b,d}^{(n_0)}(t)}{\partial t}$. In the case $n_0=1$, it thus reads
\begin{equation}
\varphi_{b,d}^{(1)}(t)=\left[e^{-dt}\varphi_0(t)+b \left(1-\int_0^{t}{\rm d}\tau\,e^{-d\tau}\varphi_0(\tau)\right) \left(\int_0^{t}{\rm d}\tau\,e^{-d\tau}\varphi_0(\tau)\right) \right]
e^{-b\int_0^{t}{\rm d}\tau\, e^{-d\tau} (t-\tau) \varphi_0(\tau)}\,.
\end{equation}

\subsubsection{Behaviour of $Q_{b,d}^{(n_0)}(t)$ at late times.} When $t\gg 1/d$ (or $t\gg 1/\mu$ if $\varphi_0(t)\sim e^{-\mu t}$ at large $t$), then $q_d(t)$ can be replaced by its asymptotic value $1-\int_{0}^{\infty}{\rm d}\tau\,e^{-d\tau}\varphi_0(\tau)= 1-\widetilde{\varphi}_0(d)$, where $\widetilde{f}(s)$ denotes the Laplace transform of $f(t)$. Similarly, the upper bound $t$ of the integral in Eq. (\ref{Q0b}) can be replaced by $\infty$ and one obtains from Eq. (\ref{Qbdn0}),
\begin{equation}
Q_{b,d}^{(n_0)}(t)\simeq c_{b,d}^{(n_0)}\,\exp(-t/\tau_{b,d}) \,,
\label{Qbdn0.large.t}
\end{equation}
where the prefactor is given by 
\begin{equation}
c_{b,d}^{(n_0)}=[1-\widetilde{\varphi}_0(d)]^{n_0} \,\exp\left[b\int_{0}^{\infty}{\rm d}\tau\ e^{-d\tau}\tau \varphi_0(\tau)\right]\,,
\label{cbd}
\end{equation}
and the characteristic reaction time by
\begin{equation}
\tau_{b,d}=\frac{1}{b\widetilde{\varphi}_0(d)}\, . 
\label{taubd}
\end{equation}
Therefore the large time decay of the no-reaction probability is exponential for any $b>0$ and the characteristic time $\tau_{b,d}$ is inversely proportional to $b$. Consequently, the MFPT is always finite. All the information about the geometry, spatial dimension and transport process is encoded into $\widetilde{\varphi}_0(d)$ for the single non-degrading particle.

\subsubsection{Mean first passage time.}\label{sec:mfptSSD} 
We denote the MFPT $\langle t_1\rangle$ as $T_{b,d}^{(n_0)}$ and use the general relation $T_{b,d}^{(n_0)}=\int_{0}^{\infty}{\rm d}t\, Q_{b,d}^{(n_0)}(t)$ to deduce
\begin{equation}
T_{b,d}^{(n_0)}=\int_0^{\infty}{\rm d}t \left[q_d(t)\right]^{n_0} e^{-b\int_0^{t}{\rm d}\tau\, e^{-d\tau} (t-\tau) \varphi_0(\tau)}\,.
\label{Tbdn0}
\end{equation}
After integrating by parts, one finds that $T_{b,d}^{(1)}$ and $T_{b,d}^{(0)}$ are simply related through
\begin{equation}
T_{b,d}^{(1)}=T_{b,d}^{(0)}-\frac{1}{b}\,.
\end{equation}
Unlike the characteristic time $\tau_{b,d}$, the MFPT does not have a simple expression and requires the knowledge of $\varphi_0(t)$ at all $t$ in principle. 

\begin{itemize}
\item {\bf Small $b$ behaviour.} In this regime the form of $T_{b,d}^{(n_0)}$ is generic: since  $q_{d}(t\to\infty)>0$ for any fixed $d>0$, then $T_{b=0,d}^{(n_0)}=\infty$ from Eq. (\ref{Tbdn0}). Therefore the  behaviour at small $b$ is controlled by the large $t$ decay of $Q_{b,d}^{(n_0)}(t)$, which is given by Eq. (\ref{Qbdn0.large.t}). Integrating this expression over the whole time domain yields,
\begin{equation}
T_{b,d}^{(n_0)}\simeq \frac{[1-\widetilde{\varphi}_0(d)]^{n_0}}{b\widetilde{\varphi}_0(d)}\to\infty\,\quad{\rm as}\ b\to 0\,,
\label{Tbdn0bsmall}
\end{equation}
which is similar to $\tau_{b,d}$ in Eq. (\ref{taubd}), but with an explicit dependence on $n_0$.

\item {\bf  Large $b$ behaviour.} In this case, $T_{b,d}^{(n_0)}$ depends on the short-time behaviour of the first passage density $\varphi_0(t)$ and can be studied in specific cases.

\item {\bf  Small $d$ behaviour.} This case does not correspond to a particularly simple limit, as the MFPT depends on the details of $\varphi_0(t)$ at all times.

\item {\bf Large $d$ behaviour.} This regime can be analysed in a way similar to the small $b$ limit above. In this case, $q_d(t)$ rapidly tends towards its asymptotic value $1-\widetilde{\varphi}_0(d)$, and Eq. (\ref{Q0a}) can be simplified to $Q_{b,d}^{(0)}(t)\simeq e^{-b\widetilde{\varphi}_0(d) t}$. The expression (\ref{Tbdn0bsmall}) is re-obtained (with $b$ fixed, not necessarily small). Since $\widetilde{\varphi}_0(d)\ll 1$ for $d$ large, we have at leading order
\begin{equation}
T_{b,d}^{(n_0)}\simeq \frac{1}{b\widetilde{\varphi}_0(d)}\to\infty\,\quad {\rm as}\ d\to\infty\,.
\label{Tbdn0dlarge}
\end{equation}

\end{itemize}
 
\subsection{Brownian target search on the semi-infinite line}\label{sec:mfptbrown}
To illustrate these results, let us revisit the example of Brownian diffusion in the one-dimensional geometry of the semi-infinite line, where an absorbing boundary is placed at the origin ($x_0>0$ in our examples).
In this case, the probability density $\varphi_0(t)$ of the first-arrival time is known explicitly and is given by the Lévy-Smirnov distribution \cite{redner2001guide}
\begin{equation}\label{4-3.33}
    \varphi_0(t) = \frac{|x_{0}|\,t^{-3/2}}{\sqrt{4\pi D}}\exp\left(
    -\frac{x_{0}^2}{4Dt}\right)\,,
\end{equation}
whose Laplace transform is $\tilde{\varphi}_0(d)=\exp[-\sqrt{\frac{d}{D}}|x_0|]$, where $D$ is the diffusion coefficient of the particle. From Eq. (\ref{taubd}), we deduce a simple and exact expression for the characteristic reaction time,
\begin{equation}
\tau_{b,d}=\frac{\e^{\sqrt{\frac{d}{D}}|x_0|}}{b}\,.
\label{taubdbrown}
\end{equation}
We observe that the rates $b$ and $d$ have highly asymmetric effects: $\tau_{b,d}$ diverges as $1/b$ when $b \to 0$, but much more rapidly than any algebraic law when $d \to \infty$.

Similarly, Eq. (\ref{Tbdn0bsmall}) can be applied to deduce the behaviour of $T_{b,d}^{(1)}$ for small $b$, yielding
\begin{equation}
T_{b,d}^{(1)}\simeq \frac{\e^{\sqrt{\frac{d}{D}}|x_0|}-1}{b}\,,\quad b\to 0\,
\label{T1bdbrownsmallb}
\end{equation}
which is similar to that of $\tau_{b,d}$. For large $d$ (and $b$ not necessarily small), we have from Eq. (\ref{Tbdn0dlarge}),
\begin{equation}
T_{b,d}^{(1)}\simeq b^{-1}\e^{\sqrt{\frac{d}{D}}|x_0|}\to\infty\,.
\end{equation}

The behaviour of $T_{b,d}^{(1)}$ or $T_{b,d}^{(n_0)}$ at large $b$ is a little more subtle to obtain and is dominated by the short $t$ regime in Eqs. (\ref{Q0b}) or (\ref{Q1}), since for non-small values of $t$, the function being integrated is an exponential with a very negative argument and turns out to be almost zero. We can then assume $[q_d(t)]^{n_0}\simeq 1$ in the prefactor. This case was studied in \cite{tung2025first} and some details are also exposed in \ref{sec:appA}. We obtain,
\begin{equation}
T_{b,d}^{(1)}\simeq \frac{x_0^2}{4D\ln\left[bx_0^2/(4D)\right]}\,,\quad b\to\infty\, .
\label{T1bdbrownlargeb}
\end{equation}
This asymptotic expression does not depend on $d$ or $n_0$. We conclude that the MFPT tends to zero as $b \to \infty$, but does so very slowly, as $1/\ln b$. Thus, for a fixed $d$, the MFPT is a monotonically decreasing function of $b$ but exhibits very different behaviours for small and large values of $b$.

\section{The cost of synthesis and its optimization}\label{sec:costs}

Keeping $d$ constant, the previous results imply that minimizing $\tau_{b,d}$ or $T_{b,d}$ with respect to $b$ leads to the optimal value $b=\infty$. However, in practice, an infinite rate cannot be achieved due to the costs associated with synthesis. 
Synthesis requires time, energy, and resources (for example, protein synthesis involves molecular machinery that uses energy from ATP to translate mRNA into proteins), for simplicity, we can assume a constant cost for each newly synthesized particle. In contrast, degradation is assumed to occur spontaneously or at a negligible cost compared with synthesis.

Therefore, a natural way to evaluate the energy cost of a search is to assume it is proportional to the total number $N_1$ of searchers employed in the interval $[0, t_1]$—that is, up to the time of first arrival. Ideally, we want the search processes to conclude rapidly but without consuming an excessive amount of resources. We construct a total cost function in which synthesis costs are added to the MFPT,
\begin{equation} \Theta_{b,d}=T^{(1)}_{b,d}+\lambda \langle N_1\rangle\,, 
\label{thetassd}
\end{equation} 
where $\lambda$ is a parameter with the dimension of time that quantifies the magnitude of the synthesis cost per particle. Here, we assume that there is one particle at $t=0$, at the start of the search process.

A simple argument allows us to obtain the average total number of particles produced during the search, $\langle N_1\rangle$. This number consists of the first particle plus the subsequent ones synthesized until the search concludes. The latter number is simply $\int_0^{T_{b,d}^{(1)}}b{\rm d}t$ —that is, the rate $b$ multiplied by the average search time $T_{b,d}^{(1)}$. Therefore,
\begin{equation}
\langle N_1\rangle=1+bT_{b,d}^{(1)}\,.
\label{n1prom}
\end{equation}
\begin{figure}[t]    
\includegraphics[width=1\textwidth]{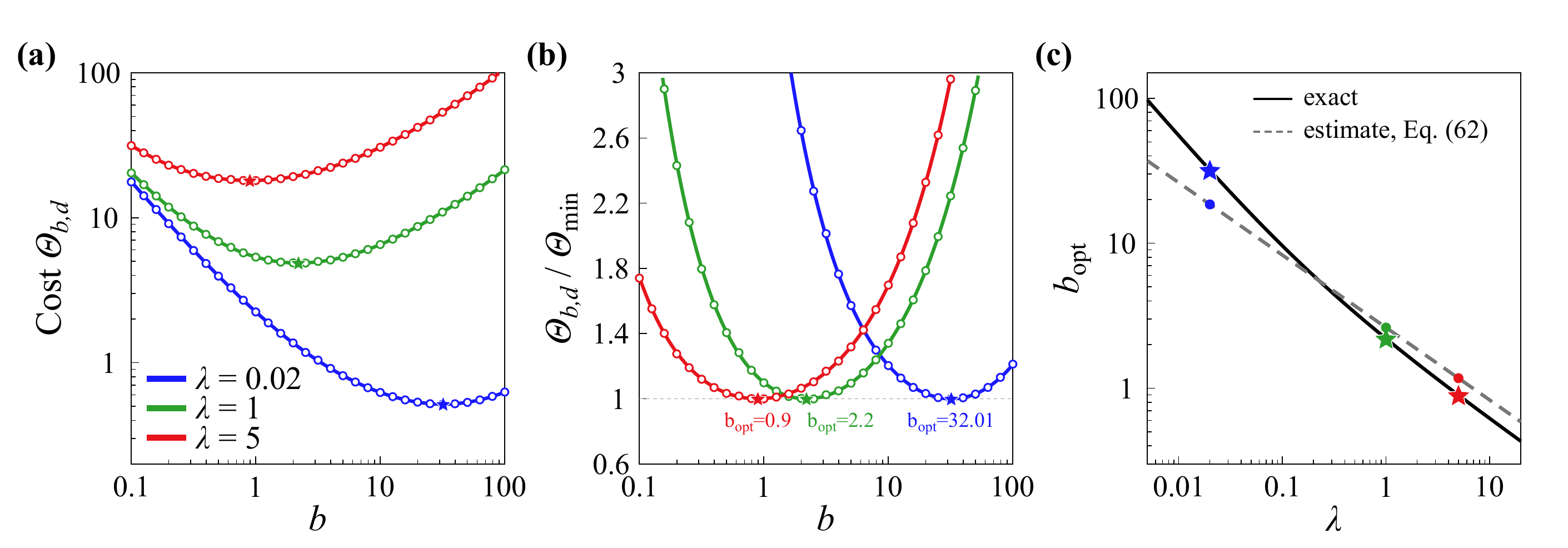}
    \centering
    \caption{(a) Total search cost as a function of the synthesis rate $b$ for different values of the synthesis cost parameter $\lambda$, with the degradation rate held constant ($d=1$). The system is composed of Brownian particles on the positive semi-infinite line and an absorbing boundary at $x=0$ ($D=1$ and $x_0=1$). (b) Same cost function rescaled by  $\Theta_{\rm min}=\underset{b\ge 0}{\min}\ \Theta_{b,d}$. (c) Optimal synthesis rate as a function of $\lambda$ and the approximation given by Eq. (\ref{bopt}). In all panels, the lines represent the numerical evaluation of the analytical expressions and the open symbols Brownian dynamics simulations of the SSD process. Each symbol represents an average over $10^6$ searches.}
    \label{fig:cost2}
\end{figure}
A slightly more elaborate argument confirms this result \cite{SSD_PRE_2026}. Combining Eqs. (\ref{thetassd}) and (\ref{n1prom}) leads us to
\begin{equation}
\Theta_{b,d}=(1+b\lambda)T^{(1)}_{b,d}+\lambda\,.
\label{thetassd.2}
\end{equation}
Now suppose that $d$ is fixed, meaning that the molecules degrade at an uncontrollable rate, while the synthesis rate is variable. In the Brownian example of the semi-infinite line, we can use Eqs. (\ref{T1bdbrownsmallb}) and (\ref{T1bdbrownlargeb}) to deduce the behaviours of $\Theta_{b,d}$ when $b\to0$ and $b\to\infty$, respectively. We obtain,
\begin{equation}
\Theta_{b,d}\simeq \frac{\e^{\sqrt{\frac{d}{D}}x_0}-1}{b}\,,\quad b\to 0\,
\label{thetassdbsmall}
\end{equation}
and
\begin{equation}
\Theta_{b,d}\simeq \frac{\lambda b x_0^2}{4D\ln\left[bx_0^2/(4D)\right]}\,,
\quad b\to \infty\,.
\label{thetassdblarge}
\end{equation}
We observe that $\Theta_{b,d}$ diverges as $1/b$ and $\lambda b/\ln b$ for small and large values of $b$, respectively. Therefore, this function reaches a global minimum at a certain finite optimal synthesis rate, $b_{\rm opt}$. Figs. \ref{fig:cost2}a-b shows the exact value of $\Theta_{b,d}$, obtained by numerically evaluating Eq. (\ref{Tbdn0}) (with fixed $d=1$) and combining it with Eq. (\ref{thetassd.2}). Results obtained from Brownian dynamics simulations (symbols) are also displayed, showing excellent agreement. The total cost exhibits a minimum for any value of $\lambda > 0$. As the synthesis cost $\lambda$ increases, the optimization becomes more sensitive to the choice of $b$ (notice the logarithmic scale on the Figure). The analytical expression for $\Theta_{b,d}$ in the intermediate range of $b$ where the minimum lies is complicated to calculate {\it a priori}. However, a naive estimate can be obtained by summing the two expressions (\ref{thetassdbsmall}) and (\ref{thetassdblarge}), differentiate with respect to $b$, and find where this derivative vanishes. Replacing the logarithmic factor with $1$ at first order, we obtain
\begin{equation}
b_{\rm opt}\approx\frac{2}{|x_0|}\sqrt{\frac{D}{\lambda}}\left(\e^{\sqrt{\frac{d}{D}}|x_0|}-1\right)^{1/2}\,.
\label{bopt}
\end{equation}
As shown by Figure \ref{fig:cost2}c, the optimal rate in Eq. (\ref{bopt}) and the value obtained from the exact numerical minimization are in reasonable agreement.

\section{SSD with bursty synthesis}\label{sec:bursts}
\begin{figure}[t]
    \centering
    \includegraphics[width=.8\textwidth]{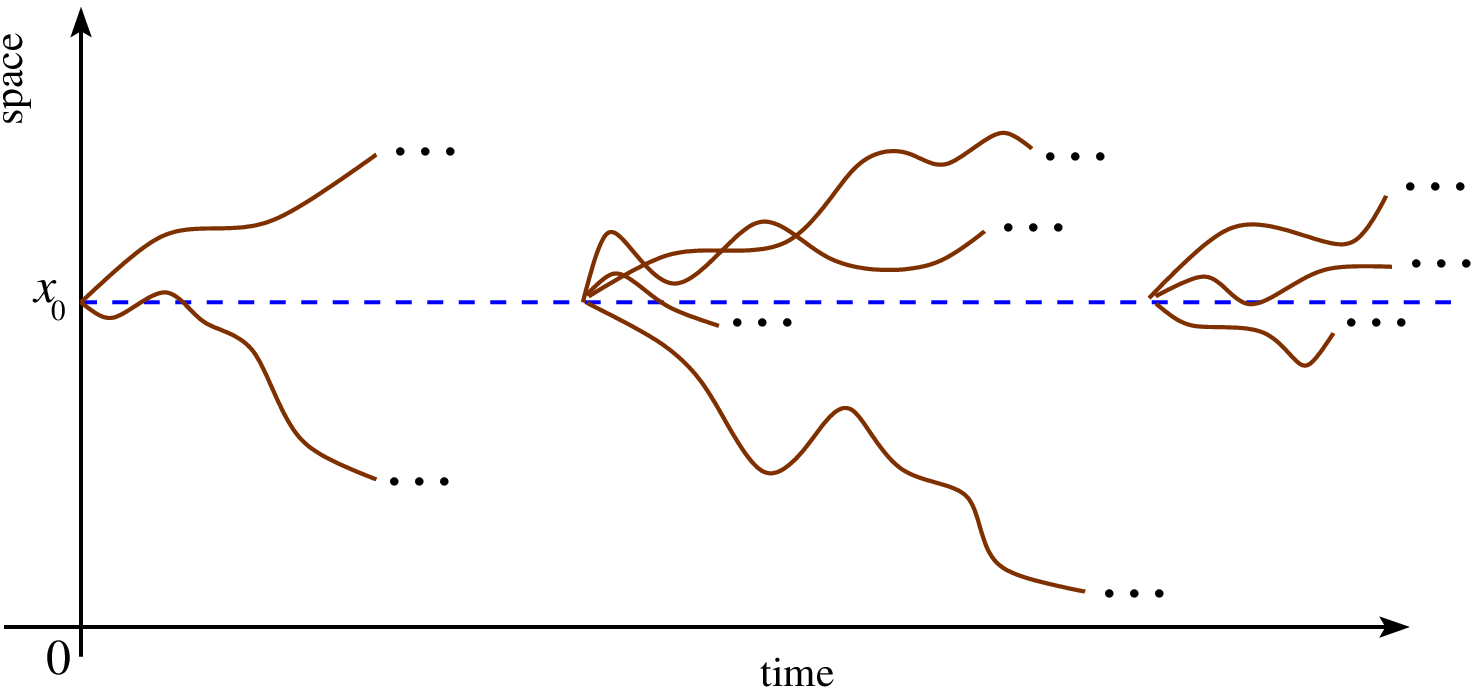}
    \caption{SSD with bursty synthesis. As in Fig. \ref{fig:N}, synthesis events occur at a rate $b$ but involve $j$ independent particles instead of $1$. The burst sizes $\{j_0,j_1,j_2,\ldots\}$ are independent random variables with distribution $\pi_j$.}
    \label{fig:burstsetup}
\end{figure}

We now consider a modification of the basic SSD model to incorporate bursty dynamics in particle production. These type of dynamics are very frequent in the cell and thought to play an important role during gene expression, in particular by causing a wide variability between cells. Protein or mRNA synthesis is often non-Poisson and does not occur unit by unit, but through events involving the production of $j$ copies within a short period of time (with $j\sim1$ to $20$, typically), separated by comparatively long periods without activity \cite{cai2006stochastic,Pedraza_2008}. Let us consider a simple model of bursty synthesis, where independent particles appear with rate $b$ in groups of size $j$ simultaneously, where $j=\{1,2,\ldots\}$ is a random integer with distribution $\pi_j$ \cite{shahrezaei2008analytical}. See Fig. \ref{fig:burstsetup} for a sketch of the process. Therefore the reactions (\ref{reaction}) become, 
\begin{equation}
\emptyset  \overset{b}{\longrightarrow} 
j\,A\quad {\rm and}\quad A \overset{d}{\longrightarrow}  \emptyset\, .
\label{reactionburst}
\end{equation}
The sizes of the successive bursts
$j_0,j_1,j_2,\ldots$ (having distribution $\pi_j$) since the initial one are supposed to be independent.  The particle number distribution $P(n)$ in the steady state can be computed in a few cases, like in the case of geometrically distributed bursts \cite{paulsson2000random}. 

\subsection{Particle density}

Let us denote $\overline{a_j}=\sum_{j=1}^{\infty}a_j\pi_j$ as the average of some quantity $a_j$ over burst sizes.
By employing the same additivity arguments as in Section \ref{sec:densgen}, the average density of independent particles with bursty synthesis at $x_0$ can be obtained by substituting $n_0$ by $\overline{j_0}=\overline{j}$ and $b{\rm d}t$ by $\overline{j} b{\rm d}t$ in Eq. (\ref{rhogen}), since $\overline {j}$ particles are produced at each synthesis on average. Hence,
\begin{equation}
\rho_{b,d}^{({\rm B})}(t)=\overline{j}\,\rho_{b,d}(t)\, ,
\label{rhoburst}
\end{equation}
where the superscript (B) denotes the bursty case and $\rho_{b,d}(t)$ is given by Eq. (\ref{rho1}) or Eq. (\ref{rhogen}) with $n_0=1$. If no particles are present in the system at $t=0$, then Eq. (\ref{rhoburst}) can also be used, setting $n_0=0$ in Eq. (\ref{rhogen}).

\subsection{No-reaction probability}\label{sec:QSSD}

The first-passage time statistics can be similarly obtained by introducing $Q_{b,d}^{({\rm B},0)}(t)$, the no-reaction probability at time $t$ of the bursty system governed by the rates $b$ and $d$, in the absence of particle at $t=0$. Similarly, $Q_{b,d}^{({\rm B},1)}(t)$ is the same probability in the presence of an initial burst at $t=0$ (and averaging over this initial burst size). The corresponding MFPTs are denoted by $T_{b,d}^{({\rm B},0)}$ and $T_{b,d}^{({\rm B},1)}$.  To compute $Q_{b,d}^{({\rm B},1)}(t)$, let us notice that $j$ non-interacting particles start to diffuse from $x_0$ at $t=0$, and averages are performed over $j$. Similarly to Eq. (\ref{Qbdn0}), we thus have
\begin{eqnarray}
Q_{b,d}^{({\rm B},1)}(t)&=&\overline{q_d(t)^j} \,Q_{b,d}^{({\rm B},0)}(t)\,\label{Q1BQ0B}\\
Q_{b,d}^{({\rm B},0)}(t)&=&e^{-bt}+b\int_0^{t}{\rm d} t'\, e^{-b(t-t')} Q_{b,d}^{({\rm B},1)}(t')\, .\label{Q0BQ1B}
\end{eqnarray}
Eq. (\ref{Q1BQ0B}) expresses the fact that $j$ independent particles are produced at once at $t=0$ (each having a no-reaction probability $q_d(t)$ at time $t$), while the subsequent synthesis events are independent from this first synthesis event. 
The renewal-like equation (\ref{Q0BQ1B}) is identical to Eq. (\ref{Q0Q1}) in the non-bursty case, recalling that the rate $b$ controls the start of the first burst in the initially empty system.
The only difference between Eqs. (\ref{Qbdn0})-(\ref{Q0Q1}) (with $n_0=1$) and Eqs. (\ref{Q1BQ0B})-(\ref{Q0BQ1B}) is the substitution  of $q_d(t)$ by $\overline{q_d(t)^j}$. We therefore immediately deduce from the solution (\ref{Q0a}) that
\begin{equation}
Q_{b,d}^{(B,0)}(t)=\exp\left[-b\int_0^{t}{\rm d}\tau\left[1-\overline{q_d(\tau)^j}\right]\right]\,,
\label{Q0B}
\end{equation}
and
\begin{equation}
Q_{b,d}^{(B,1)}(t)= \overline{q_d(t)^j}\,
\exp\left[-b\int_0^{t}{\rm d}\tau\left[1-\overline{q_d(\tau)^j}\right]\right]\,.
\label{Q1B}
\end{equation}
At large time $t$, it is easy to see from Eqs. (\ref{Q0B})-(\ref{Q1B}) that
\begin{equation}
Q_{b,d}^{(B,0)}(t)\propto Q_{b,d}^{(B,1)}\propto e^{-t/\tau_{b,d}^{(B)}}
\end{equation}
with a characteristic reaction time,
\begin{equation}
\tau_{b,d}^{(B)}=\frac{1}{b\left[1-\overline{q_d(\infty)^j}\right]}\, ,
\end{equation}
where $q_d(\infty)=1-\widetilde{\varphi}_0(d)$.
This expression is valid for any $b$ and $d$.

\subsection{Mean first-passage times}

We use the previous results to calculate the MFPT $T_{b,d}^{(B,1)}=\int_0^{\infty}{\rm d}t\, Q_{b,d}^{(B,1)}(t)$ and $T_{b,d}^{(B,0)}=\int_0^{\infty}{\rm d}t\, Q_{b,d}^{(B,0)}(t)$.

\begin{itemize}
\item {\bf Small $b$ behaviour.} Once again, the small $b$ behaviours of the MFPTs follow from substituting the large $t$ behaviours of the no-reaction probabilities over the whole temporal range in the integrals $T_{b,d}^{(B,1)}=\int_0^{\infty}{\rm d}t\, Q_{b,d}^{(B,1)}(t)$  and $T_{b,d}^{(B,0)}=\int_0^{\infty}{\rm d}t\, Q_{b,d}^{(B,0)}(t)$. For small $b$, $Q_{b,d}^{(B,0)}(t)\simeq e^{-t/\tau_{b,d}^{(B)}}$ from Eq. (\ref{Q0B}) (as the prefactor in front of the exponential is close to unity) and one obtains
\begin{equation}
T_{b,d}^{(B,0)}\simeq \frac{1}{b\left[1-\overline{q_d(\infty)^j}\right]}\,\quad {\rm and} \quad
T_{b,d}^{(B,1)}\simeq \frac{\overline{q_d(\infty)^j}}{b\left[1-\overline{q_d(\infty)^j}\right]}\,,\quad b\to 0\,.
\label{TbdBsmallb}
\end{equation}
When $\pi_j=\delta_{j,1}$, Eq. (\ref{Tbdn0bsmall}) with $n_0=0$ and $n_0=1$ is recovered, respectively.

\item {\bf Large $b$ behaviour.} This regime is controlled by the small $t$ regime of Eqs. (\ref{Q0B})-(\ref{Q1B}), where $q_d(t)$ and $q_d(\tau)$ are close to 1. Writing $q_d(\tau)\equiv 1-c_d(\tau)$ with $c_d(\tau)\ll 1$ and using $\overline{q_d(\tau)^j}\simeq 1-\overline{j} c_d(\tau)$ at leading order, one obtains 
\begin{equation}
T_{b,d}^{(B,0)}\simeq T_{b,d}^{(B,1)}\simeq\int_0^{\infty}{\rm d}t\,
e^{-b\int_0^t{\rm d}\tau\, \overline{j}\,[1-q_d(\tau)]}\,.
\end{equation}
One deduces
\begin{equation}
T_{b,d}^{(B,1)}\simeq T_{\overline{j} b,d}^{(1)}\,,\quad b\to\infty\,.
\label{TbdBlargeb}
\end{equation}
Hence, for a process with mean burst size $\overline{j}$, the MFPT becomes equal to the MFPT of a process producing one particle at a time, but at a rate $\overline{j}$ times as fast. We stress that this equivalence only holds in the limit $b\to\infty$.

Unlike the small $b$ expansion (\ref{TbdBsmallb}), where the MFPT is always inversely proportional to $b$, we saw that the explicit behaviour of $T_{b,d}^{(1)}$ at large $b$ depends on the underlying particle motion. For 1D Brownian motion on the semi-infinite line with an absorbing boundary located at the origin, studied in Section \ref{sec:mfptbrown}, one can combine Eq. (\ref{TbdBlargeb}) and Eq. (\ref{T1bdbrownlargeb}) to deduce
\begin{equation}
T_{b,d}^{(B,1)}\simeq \frac{x_0^2}{4D\ln\left[\overline{j}b x_0^2/(4D)\right]}\,,\quad b\to\infty.
\label{TbdBlargebBrown}
\end{equation}
Since the average burst size only enters in the logarithm, for very large synthesis rate the MFPT is little affected by the bursty character of the dynamics. Hence, when $\overline{j}$ is sufficiently large, increasing its value does not lead to a substantial gain in the MFPT, contrary to the following case.

\item {\bf Large $d$ behaviour}. This regime can be analysed by applying the same arguments as exposed in section \ref{sec:mfptSSD}, where $q_d(\infty)$ is substituted by $\overline{q_d(\infty)^j}$. If $d$ is large then $\widetilde{\varphi}_0(d)\ll 1$, therefore $\overline{q_d(\infty)^j}\simeq 1-\overline{j}\widetilde{\varphi}_0(d)$. Eq. (\ref{Tbdn0dlarge}) is generalized to
\begin{equation}
T_{b,d}^{(B,1)}\simeq \frac{1}{\overline{j}b\widetilde{\varphi}_0(d)}\,,\quad d\to\infty\,.
\label{TbdBlarged}
\end{equation}
Hence the MFPT is inversely proportional to $b$ and to the mean burst size in the limit $d\to\infty$.
\end{itemize}

\subsection{Comparison between different types of synthesis.} It is instructive to compare the MFPT in three settings: $(i)$ burst sizes are distributed with an arbitrary distribution $\pi_j$ of mean $\overline{j}\ge 1$, $(ii)$ all bursts have the same size $\overline{j}$, i.e., with distribution $\delta_{j,\overline{j}}$, $(iii)$ the bursts are of size 1 and occur with a rate $\overline{j}b$. For a fair comparison let us assume $n_0=0$ initially, so that the mean number of particles  is the same at all times in these three cases, according to Eq. (\ref{meannb}) with $b$ replaced by $\overline{j}b$. 

Given the concavity of $x^j$ with respect to the variable $j$, we have $\overline{x^j}\ge x^{\overline{j}}$, therefore  $-b[1-\overline{q_d(t)^j}]\ge -b[1-q_d(t)^{\overline{j}}]$, leading to the inequalities $Q_{b,d}^{(B,0)}[\pi_j](t)\ge Q_{b,d}^{(B,0)}[\delta_{j,\overline{j}}](t)$ and    $T_{b,d}^{(B,0)}[\pi_j]\ge T_{b,d}^{(B,0)}[\delta_{j,\overline{j}}]$.

The lower bound $T_{b,d}^{(B,0)}[\delta_{j,\overline{j}}]$ can be further compared to $T_{\overline{j}b,d}^{(0)}$. Let us define the functions $f(x)=-(1-x^{\overline{j}})$ and $g(x)=-\overline{j}(1-x)$. Since $f(1)=g(1)=0$, $f'(1)=g'(1)=\overline{j}$, and $f''(x)\ge 0$ while $g''(x)=0$, then $f(x)\ge g(x)$ for all $x\ge 0$. Using this relation in  Eqs. (\ref{Q0a}) and (\ref{Q0B}), yields  $Q_{b,d}^{(B,0)}[\delta_{j,\overline{j}}](t)\ge Q_{\overline{j}b,d}^{(0)}(t)$ and similarly for the MFPTs. In summary, we have
\begin{equation}
T_{b,d}^{(B,0)}[\pi_j]\ge T_{b,d}^{(B,0)}[\delta_{j,\overline{j}}]\ge T_{\overline{j}b,d}^{(0)}\,.
\label{inequalities}
\end{equation}
\begin{figure}[t]
    \centering
    \includegraphics[width=.45\textwidth,angle=-90]{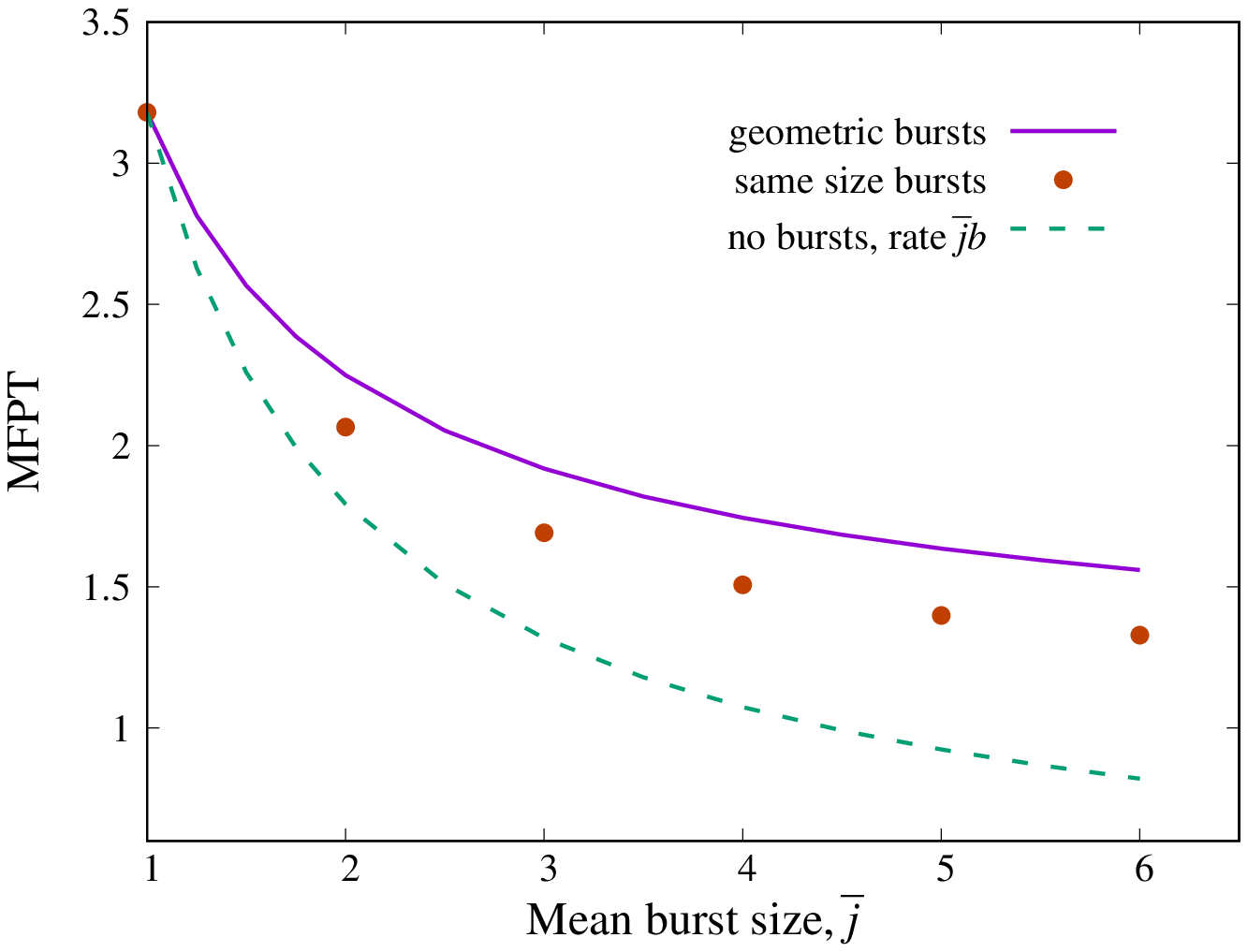}
    \caption{Comparison between different synthesis processes with $n_0=0$ initial particles, for Brownian searchers on the semi-infinite line and an absorbing boundary at $0$ (other parameters are $b=d=1$, $x_0=1$ and $D=1$). Bursts of mean size $\overline{j}$ and distributed geometrically [Eq. (\ref{geodist})] lead to longer MFPTs than bursts of equal size 
    $\overline{j}$. Ordinary SSD with a synthesis rate $\overline{j}b$ achieves even lower MFPTs.}
    \label{fig:comp}
\end{figure}
Hence, an SSD process producing one particle at a time at a high rate $\overline{j}b$ is comparatively more efficient for search than a process producing $\overline{j}$ particles at once with a rate $b$, which is itself more efficient than the SSD process producing unequal bursts of mean size $\overline{j}$.  
We conclude that increasing fluctuations in the synthesis dynamics prolong the reaction with the absorbing target site. The inequalities (\ref{inequalities}) are verified in Fig. \ref{fig:comp}, that shows the variations of the different MFPT with respect to $\overline{j}$ in the case of Brownian particles produced at $x_0>0$ on the semi-infinite line, with an absorbing boundary at $x=0$. In the upper curve, the burst sizes are distributed geometrically (see section \ref{sec:geo} below).  

\subsection{Geometric distribution of burst sizes.}\label{sec:geo} 

In many experimental observations, the burst size approximately follows a geometric (or exponential) distribution \cite{cai2006stochastic,shahrezaei2008analytical}. 
It is given by
\begin{equation}
\pi_j=\frac{1}{\overline{j}-1}\left(\frac{\overline{j}-1}{\overline{j}}\right)^j\,,\quad j=1,2,\ldots\,,
\label{geodist}
\end{equation}
where $\overline{j}=\sum_{j=1}^{\infty}j\pi_j\ge 1$ is a free parameter. Inserting Eq. (\ref{geodist}) into Eqs. (\ref{Q0B})-(\ref{Q1B}) and integrating over $t$, one obtains the exact expressions
\begin{eqnarray}
\fl
T_{b,d}^{(B,0)}&=&\int_{0}^{\infty}{\rm d}t\, \exp\left[-\overline{j}b\int_0^{t}{\rm d}\tau\,\frac{1-q_d(\tau)}{\overline{j}[1-q_d(\tau)]+q_d(\tau)}\right]\,,\label{T0Bgeo}\\
\fl
T_{b,d}^{(B,1)}&=&\int_{0}^{\infty}{\rm d}t\, \frac{q_d(\tau)}{\overline{j}[1-q_d(\tau)]+q_d(\tau)}\,\exp\left[-\overline{j}b\int_0^{t}{\rm d}\tau\,\frac{1-q_d(\tau)}{\overline{j}[1-q_d(\tau)]+q_d(\tau)}\right]\,.\label{T1Bgeo}
\end{eqnarray}
\begin{figure}[t]
    \centering
    \includegraphics[width=.34\textwidth,angle=-90]{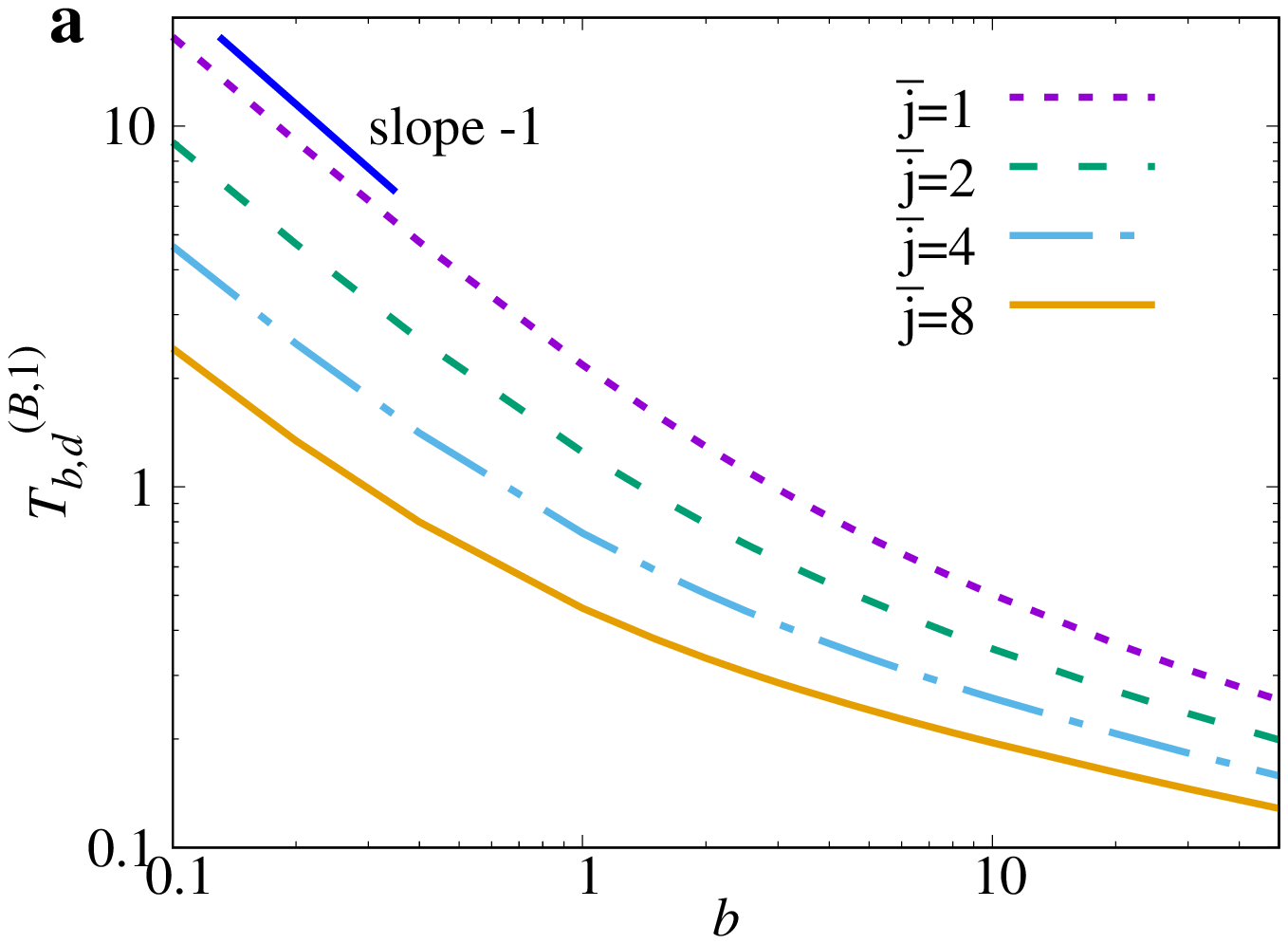}\hspace{0.3cm}\includegraphics[width=.34\textwidth,angle=-90]{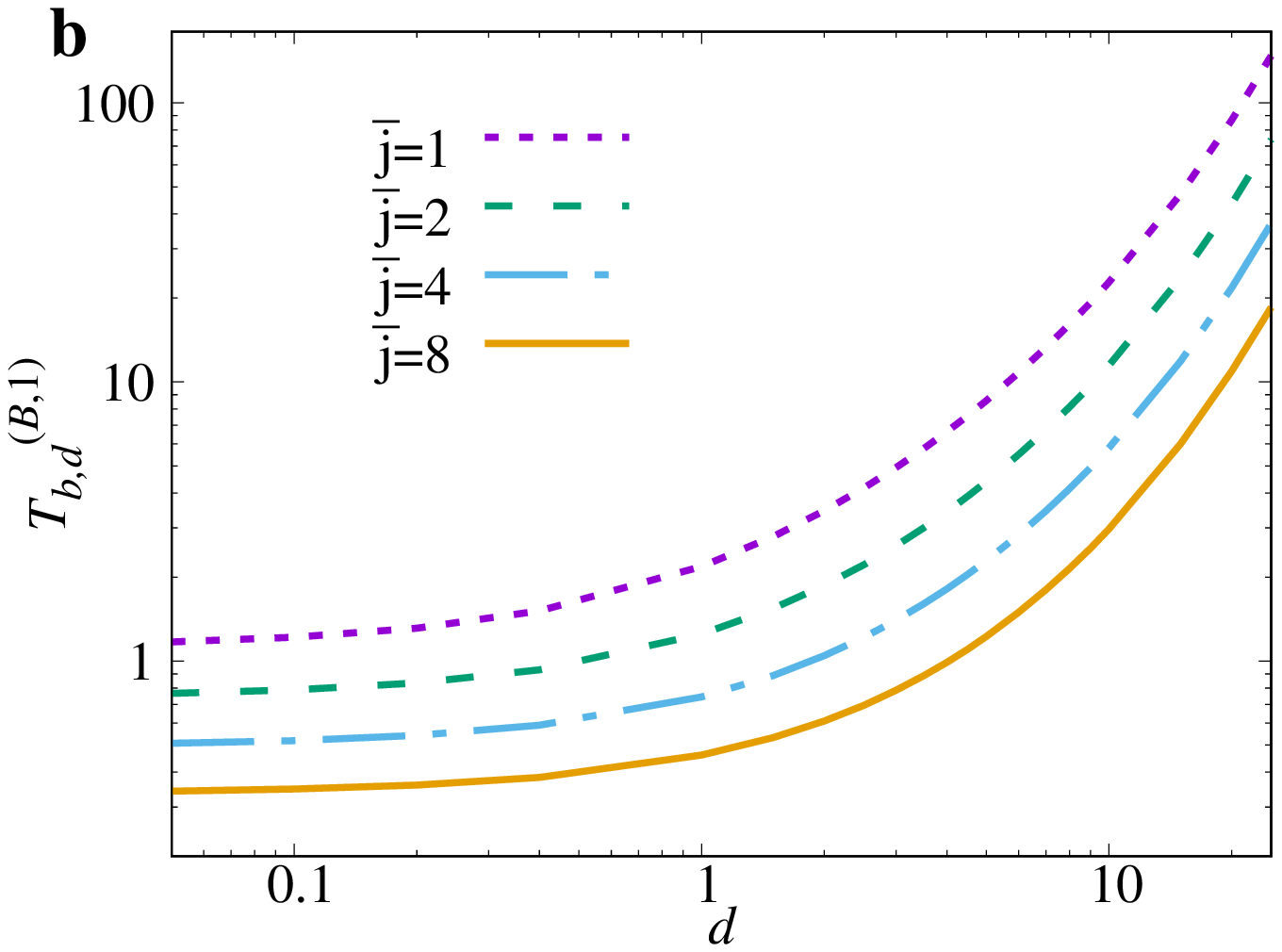}

    \includegraphics[width=.34\textwidth,angle=-90]{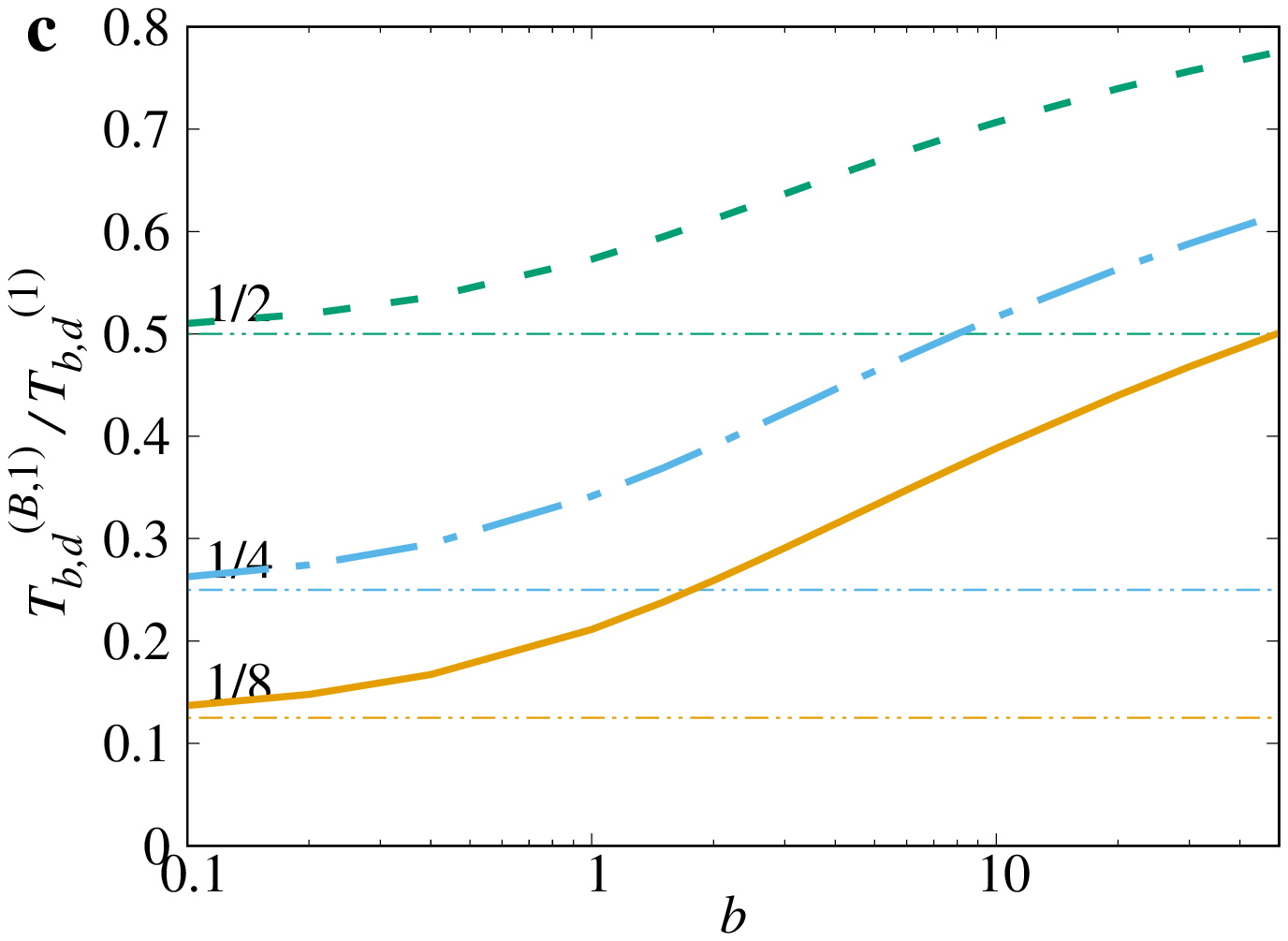}\hspace{0.3cm}
    \includegraphics[width=.34\textwidth,angle=-90]{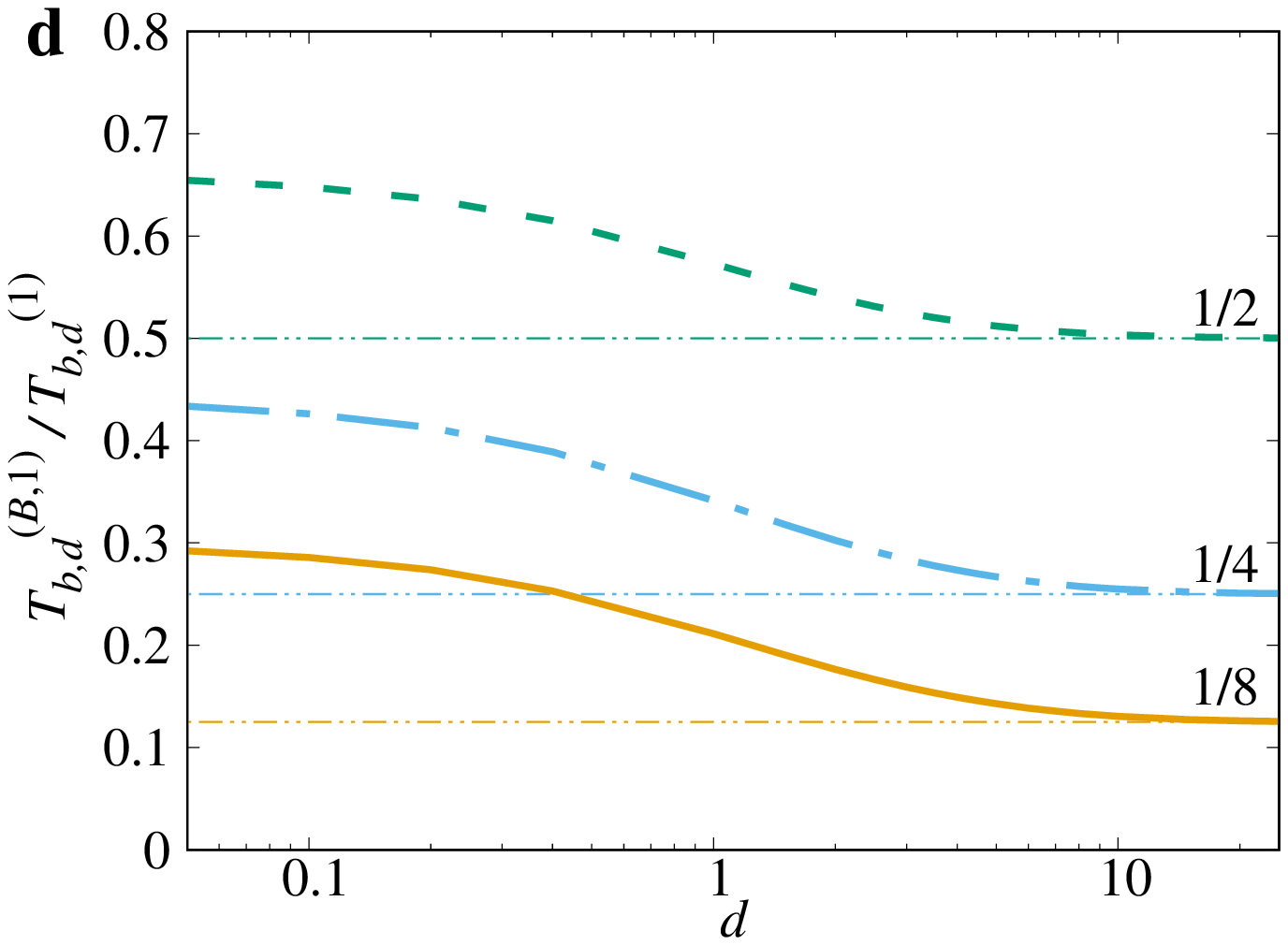}
    \caption{(a) MFPT with one initial burst as a function of the burst synthesis rate $b$ for several mean burst sizes $\overline{j}$ ($d=1$ and $j$ is geometrically distributed). The other parameters are the same as in Fig. \ref{fig:comp}. (b) Same quantity as a function of the degradation rate $d$ (while $b=1$). (c) Relative gain in the search time shown in (a) as a function of $b$ (same colour code), which tends to $1/\overline{j}$ at small $b$.
    (d) Relative gain in the mean search time shown in (b) as a function of $d$, which also tends to $1/\overline{j}$ at large $d$.}
    \label{fig:TB1}
\end{figure}
The small $b$ regime in Eq. (\ref{TbdBsmallb}) can be explicitly calculated for all $d$ as,
\begin{equation}
T_{b,d}^{(B,1)}\simeq \frac{1-\widetilde{\varphi}_0(d)}{\overline{j}b\widetilde{\varphi}_0(d)}\,,\quad b\to 0\,.
\label{TbdB1bsmallgeo}
\end{equation}

A conclusion reached by Eqs. (\ref{TbdBlarged}) (in general) and (\ref{TbdB1bsmallgeo}) (for the geometric distribution) is the following: when the search process carried out by the standard SSD process becomes very long, either because $b$ is small or $d$ is large, then the MFPT can be reduced by a factor $\overline{j}$ thanks to bursty dynamics. In other words, bursty synthesis provides an efficient strategy to mitigate long searching times in unfavourable conditions.

We illustrate these findings through the previous one-dimensional Brownian example. Fig. \ref{fig:TB1}a displays $T_{b,d}^{(B,1)}$ as a function of $b$ for several mean burst sizes $\overline{j}$, where the $1/b$ scaling at small $b$ predicted by Eq. (\ref{TbdBsmallb}) or (\ref{TbdB1bsmallgeo}) is observed. At large $b$, the relative variations between the different curves are less pronounced, as expected from Eq. (\ref{TbdBlargebBrown}). The $1/\overline{j}$ scaling in Eq. (\ref{TbdB1bsmallgeo}) in the small $b$ limit is verified in Fig. \ref{fig:TB1}c, through the quantity $T_{b,d}^{(B,1)}/T_{b,d}^{(1)}$, for several values of $\overline{j}$.

Conversely, the increasing behaviour of $T_{b,d}^{(B,1)}$ with $d$ is shown in Fig.
\ref{fig:TB1}b, while Fig.
\ref{fig:TB1}d provides a check of the $1/\overline{j}$ factor that shows up at large $d$ in the MFPT, as predicted by Eq. (\ref{TbdBlarged}).

\section{Conclusions}\label{sec:concl}

We have studied the density profiles and first passage times of a system composed of non-interacting stochastic processes that are introduced and removed at random times with rates $b$ and $d$, respectively. The distribution of the number of particles present in the system at a given time is well studied in queueing theory, and follows a $M/M/\infty$ queue. The SSD model studied here is relevant, for instance, to describe the stochastic synthesis of proteins that diffuse in the cell and whose individual fate is either to react with a target site or to be degraded before completing their function. In another context recently discussed, the processes can be stock prices of firms that enter a market one by one at a given initial price, evolve and then exit the market at a later time \cite{pal2026geometric}. SSD processes can also model population dynamics with immigration, where the individuals enter the system from a source \cite{campos2024dynamics,tung2025first}, with potential consequences for epidemic spreading \cite{gutierrez2026heterogeneous}.

The density of particles exhibits at large times a non-equilibrium steady state, see Eq. (\ref{rhoness}), which is expressed quite generally in terms of the propagator of a single particle without degradation and starting at $x_0$, the position of the source. Except for a prefactor given by the mean number of particles $b/d$, this expression is identical to the NESS of a single particle subject to stochastic resetting to $x_0$ with rate $d$.
With a particular initial condition, the two densities are even proportional to each other at all times, for any underlying process followed by the single particle. However, the probabilistic interpretation of the density in SSD is completely different than in resetting. We also emphasize that this connection holds when the resetting and synthesis/degradation processes are Markov, i.e., occur at constant rates. Consequently, the dynamical transition in the temporal relaxation toward the steady state observed in simple diffusion under resetting \cite{majumdar2015dynamical}, also takes place in SSD.

When the particle synthesis is non-Markov, i.e., $b$ not constant in time, characterizing the SSD density becomes more involved and deserves further study. In principle, the aforementioned relation with resetting breaks down, and the density in SSD has no reason to match that of a process with a time-dependent resetting rate \cite{pal_diffusion_2016}.

We have also studied the efficiency of the search processes carried out by SSD processes.
In this case as well, it is possible to express in a general way the no-reaction probability (the probability that no particle has hit a prescribed target site up to time $t$) in terms of the first passage properties of a single particle without degradation. This relation is valid for any dimensionality, geometry or type of motion followed by the individual particles. While a process with $b=0$ and $d=0$ may take an infinite mean time to react with a target site, the MFPT always becomes finite for any $b>0$, even in the presence of degradation ($d>0$). At large times, the decay of the no-reaction probability is exponential with a characteristic time given by Eq. (\ref{taubd}). The small $b$ behaviour of the MFPT is relatively simple, see Eq. (\ref{Tbdn0bsmall}), but the large $b$ limit depends on the specific short-time behaviour of the first-passage density of the underlying process. 

In SSD searches, the synthesis and degradation rates have antagonistic effects on the mean search time, which is a decreasing function of the former and increasing function of the latter. This property has an impact on the total search costs and leads to an optimal synthesis rate (at fixed degradation) when the synthesis costs are taken into account.

We have generalized the results to bursty synthesis, when each production event involves the release of $j\ge 1$ particles at the same time instead of one, a phenomenon often observed in cell biology. When standard synthesis leads to long search times (i.e., when $b$ is small or $d$ is large), the bursts allow to divide the MFPT by the mean burst size $\bar{j}$, independently of the underlying process. This factor is significantly larger than  $\ln \bar{j}$, the factor which usually appears in Brownian searches with $N=\bar{j}$ initial non-degrading particles and without further injection of elements \cite{lawley2020distribution}.  

It would be interesting to study extensions of the basic model, for instance when the synthesis points are distributed in space, a situation which strongly impacts the search times in the case of searchers launched simultaneously \cite{grebenkov2020single}, or when the reaction with the target site is imperfect. The case of interacting processes also deserves further study.

\ack 
This research was supported in part by grants from the NSF (DMS-2235451) and Simons Foundation (MPS-NITMB-00005320) to the NSF-Simons National Institute for Theory and Mathematics in Biology (NITMB). We thanks Samantha Linn, Aanjaneya Kumar and Pedro Mart\'\i nez Lim\'on for useful discussions.

\appendix
\section{Large $b$ behaviour of the MFPT for Brownian particles}\label{sec:appA}
In this limit one can set $d=0$, as the search typically ends before the particles have degraded. Taking $n_0=0$ for simplicity, the no-reaction probability at time $t$ can be written as 
\begin{equation}
S_{b,d}^{(0)}(t)=\e^{-bf(t)}\,,\quad f(t)=\int_0^t{\rm d}\tau\, g(\tau)\,,\quad g(\tau)=\int_0^{\tau}{\rm d}u\,\varphi_0(u)\,.
\label{ap13.fdef}
\end{equation}
If $\varphi_0(u)$ is the Lévy-Smirnov distribution in Eq. (\ref{4-3.33}), then
\begin{equation}
g(\tau)=\frac{x_0}{\sqrt{4\pi D}}\int_0^{\tau}{\rm d}u\,\frac{1}{u^{3/2}}\e^{-\frac{x_0^2}{4Du}}=\sqrt{\frac{4D}{\pi}}\frac{1}{x_0}\int_0^{\tau}{\rm d}u\, u^{1/2}\frac{\rm d}{{\rm d}u}\left[\e^{-\frac{x_0^2}{4Du}} \right]\,.
\label{ap13.g0}
\end{equation}
Integrating by parts,
\begin{equation}
g(\tau)=\sqrt{\frac{4D}{\pi}}\frac{1}{x_0}\tau^{1/2}\e^{-\frac{x_0^2}{4D\tau}}-\sqrt{\frac{D}{\pi}}\frac{1}{x_0}\int_0^{\tau}{\rm d}u\,\frac{1}{u^{1/2}}\e^{-\frac{x_0^2}{4Du}}\,.
\label{ap13.g1}
\end{equation}
At large $b$, the MFPT is dominated by the behaviour of $\varphi_0(t)$ at small $t$, hence the variables $\tau$ and $u$ are small, and $1/u^{1/2} \ll 1/u^{3/2}$. Therefore, the second term on the right-hand side of Eq.~\eqref{ap13.g1} is much smaller than $g(\tau)$ (in the first equality in Eq.~\eqref{ap13.g0}) and  negligible. Substituting into the definition of $f(t)$—i.e., into Eq.~\eqref{ap13.fdef}—we obtain
\begin{equation}
f(t)\simeq \sqrt{\frac{4D}{\pi}}\frac{1}{x_0}\int_0^t {\rm d}\tau\, \tau^{1/2}\e^{-\frac{x_0^2}{4D\tau}}=\left(\frac{4D}{x_0^2}\right)^{3/2}\frac{1}{\sqrt{\pi}} 
\int_0^t {\rm d}\tau\, \tau^{5/2}\frac{\rm d}{{\rm d}\tau}\left[\e^{-\frac{x_0^2}{4D\tau}}\right]\,.
\label{ap13.f1}
\end{equation}
Integrating again by parts,
\begin{equation}
f(t)\simeq \left(\frac{4D}{x_0^2}\right)^{3/2}\frac{1}{\sqrt{\pi}}\,t^{5/2}\e^{-\frac{x_0^2}{4Dt}}\,+\,\ldots\,,
\label{ap13.f2}
\end{equation}
where $\ldots$ is again a negligible integral, as it involves an integral with a factor $\tau^{3/2}$ which is $\ll \tau^{1/2}$; see the first equality in Eq.~\eqref{ap13.f1}. When $b$ is large, the MFPT can be obtained by using the approximation for small $t$ extended to the entire time domain,
\begin{equation}
T_{b,d}^{(0)}\simeq T_{b,d}^{(1)}\simeq \int_0^{\infty}{\rm d}t\, \e^{-bf(t)}\,,
\end{equation}
with $f(t)$ given by Eq.~\eqref{ap13.f2}. Due to its $\e^{-c/t}$ behaviour at small $t$, the function $f(t)$ is practically zero for small $t$, up to a value $t^*$ such that $bf(t^*)=a$, where $a$ is a constant of order $1$. Then
\begin{equation}
T_{b,d}^{(1)}\simeq \int_0^{t^*}{\rm d}t\,\e^{-0}\,+\,\int_{t^*}^{\infty}{\rm d}t\ \exp\left[-b \left(\frac{4D}{x_0^2}\right)^{3/2}\frac{1}{\sqrt{\pi}}\,t^{5/2}\e^{-\frac{x_0^2}{4Dt}}\right]\simeq t^*\,,
\label{ap13.t1}
\end{equation}
where the second integral on the right-hand side of Eq.~\eqref{ap13.t1} has been neglected because it is of order $b^{-2/5}$ and much smaller than $t^*$, which is of order $1/\ln b$. The solution of $bf(t^*)=a$ is obtained from Eq.~\eqref{ap13.f2} to first order as
\begin{equation}
t^*\simeq\frac{x_0^2}{4D\ln\left[b x_0^2/(4Da\sqrt{\pi})\right]}\,. 
\label{ap13.tstar}
\end{equation}
The value of $a$ does not matter at leading order, given the presence of a large $b$ in the argument of the logarithm; thus, we can choose $a\sqrt{\pi}=1$. We retain the other constants to ensure that the argument of the logarithm is dimensionless. Combining Eqs.~\eqref{ap13.t1} and \eqref{ap13.tstar} yields Eq.~\eqref{T1bdbrownlargeb} from the main text.

\section*{References}
\bibliographystyle{iopart-num-long}
\providecommand{\newblock}{}

\end{document}